\documentclass[prd,aps,floats,floatfix,superscriptaddress,preprintnumbers,
showpacs,eqsecnum,nofootinbib,twocolumn]{revtex4}
\usepackage{amsfonts,amsmath,amssymb,latexsym,array,
theorem,mathrsfs,subfigure,bm,float,color,graphicx}

\definecolor{red}{rgb}{1,0,0}
\definecolor{blue}{rgb}{0,0,1}
\definecolor{green}{rgb}{0,1,0}

\DeclareMathAlphabet{\mathpzc}{OT1}{pzc}{m}{it}

\newcommand{\ma}[1]{\mbox{$\mathcal{#1}$}}

\newcommand{\red}[1]{{{\textcolor{red}{#1}}}}
\newcommand{\blue}[1]{{{\textcolor{blue}{#1}}}}
\newcommand{\green}[1]{{{\textcolor{green}{#1}}}}

\newcommand{\calhR}[1]{\raisebox{2ex}{\tiny ({\em h})}\hspace{-0.8em}{\ma R}}

\begin{document}

\title{
Oscillating Universe in Ho\v{r}ava-Lifshitz Gravity
}

\author{Kei-ichi {\sc Maeda}}
\email{maeda@waseda.jp}
\address{Department of Physics, Waseda University, 
Okubo 3-4-1, Shinjuku, Tokyo 169-8555, Japan}
\address{RISE, Waseda University,
Okubo 3-4-1, Shinjuku, Tokyo 169-8555, Japan}
\author{Yosuke {\sc Misonoh}}
\email{y"underscore"misonou@moegi.waseda.jp}
\address{Department of Physics, Waseda University, 
Okubo 3-4-1, Shinjuku, Tokyo 169-8555, Japan}
\author{Tsutomu {\sc Kobayashi}}
\email{tsutomu@resceu.s.u-tokyo.ac.jp}
\address{Research Center for the Early Universe (RESCEU),
Graduate School of Science, \\
The University of Tokyo, Tokyo 113-0033, Japan}

\date{\today}

\begin{abstract}
We study the dynamics of
isotropic and homogeneous universes 
in the generalized Ho\v{r}ava-Lifshitz gravity,
 and classify all possible evolutions of
vacuum spacetime.
In the case without
the detailed balance condition, we find a variety of phase structures
 of vacuum spacetimes depending on the coupling constants 
as well as the spatial curvature $K$ and a cosmological constant
 $\Lambda$.
A bounce universe solution is obtained for  $\Lambda> 0, K=\pm 1$ or 
 $\Lambda= 0, K=- 1$,
while an oscillation spacetime is found for $\Lambda\geq  0, K=1$,
or  $\Lambda< 0, K=\pm 1$.
We also propose a quantum tunneling scenario from
an oscillating spacetime to an inflationary universe,
resulting in a macroscopic cyclic universe.

\end{abstract}

\pacs{04.60.-m, 98.80.Cq, 98.80.-k} 

\maketitle

\section{Introduction}
\label{introdunction}

Recently Ho\v{r}ava proposed a power-counting renormalizable theory 
of gravity~\cite{Horava},
which has attracted much attention over the past year.
In Ho\v{r}ava's theory, Lorentz symmetry is broken and
it exhibits a Lifshitz-like anisotropic scaling in the ultraviolet (UV),
$t\to\ell^z t, \Vec{x}\to\ell\Vec{x}$,
with the dynamical critical exponent $z = 3$.
(For this reason the theory is called Ho\v{r}ava-Lifshitz (HL) gravity.)
It is then natural to expect that
the UV behavior of the theory would give rise to new scenarios of
cosmology~\cite{Calc, Kiritsis,Brandenberger}.
Earlier works have indeed revealed some interesting
aspects of HL cosmology such as
dark matter as integration constant~\cite{DM},
the generation of chiral gravitational waves from 
inflation~\cite{Takahashi-Soda},
scale-invariant fluctuations without inflation~\cite{scinv}, 
and possible dark energy scenario 
\cite{cosmology6,cosmology11,cosmology14}.
There are also some discussion 
closely related to  observational cosmology and astrophysics
such as
cosmological perturbation~\cite{scinv,cosmology1,cosmology2,cosmology4,
cosmology10,cosmology52,WangM,cosmology12,
cosmology13,cosmology18,cosmology22,cosmology27,addition01},
observational constraints~\cite{cosmology3,cosmology9,cosmology19},
primordial magnetic field without inflation~\cite{cosmology15}, 
and a relativistic star~\cite{Izumi,Greenwald}.

Though the viability of HL gravity is still under intense debate
\cite{addition02,Charm,Li,addition03,BPS_1,K-A,Henn,DM,DM2,BPS_2,Papazoglou:2009fj,BPS_3,KP,addition04},
we give the theory the benefit of the doubt
and will furthermore pursue consequences of Ho\v{r}ava's intriguing idea.

Here we focus on the dynamics of 
Friedmann-Lemaitre-Robertson-Walker (FLRW)
 universe in HL gravity, which may
provide us new aspect of the early universe. 
We study  the 
 FLRW spacetime
with {\em non-zero} spatial curvature in the context of HL gravity.
The non-trivial cosmological evolution is brought by
the various terms in the potential which are
constructed from the spatial curvature $\mathpzc{R}_{\,ij}$.
In particular, we find non-singular behavior in high curvature 
region, which may lead to avoidance of a big bang initial singularity
\cite{Brandenberger}.
Consequently, many studies on the dynamics of the 
 FLRW universe
 in HL gravity come out within the last few years
\cite{cosmology36,Minamitsuji:2009ii,Wang:2009rw,cosmology38,
cosmology16,cosmology17,cosmology39,cosmology25,cosmology50,addition05,
cosmology5,Carloni:2009jc,
cosmology31,cosmology32,cosmology33,emergent_universe,cosmology35,Calc,Brandenberger,cosmology23,
Kiritsis,CaiSar}.

In the original HL gravity,
the so-called detailed balance condition is assumed.
However, this condition can be loosened
to have arbitrary coupling constants \cite{SVW}.
Hence both models have been so far 
discussed in the analysis of the  FLRW universe.
As for matter components, a perfect fluid with the equation of state 
$P = w \rho$ and a scalar field have been discussed.
As a result, we can classify 
the analyzed models into four types:
(1) The model with the detailed balance condition and
with a perfect fluid
~\cite{cosmology36,Minamitsuji:2009ii,Wang:2009rw,cosmology38,
cosmology16,cosmology17,cosmology39,cosmology25,cosmology50,addition05}, (2) 
 The model with the detailed balance condition  and 
with a scalar field~\cite{Calc,Brandenberger,cosmology23}
(3) The model without the detailed balance condition  and
with a perfect fluid
~\cite{cosmology5,Carloni:2009jc,
cosmology31,cosmology32,cosmology33,emergent_universe,cosmology35}, and (4) 
The model without the detailed balance condition  and with 
a scalar field~\cite{Kiritsis,CaiSar}. 

So far many works have been done assuming the detailed
balance condition, and they reveal
the possibility of singularity avoidance such as a bounce 
universe \cite{Calc,Kiritsis,Brandenberger,cosmology6,cosmology22,
Minamitsuji:2009ii,Wang:2009rw,cosmology23} and an oscillating spacetime 
\cite{Minamitsuji:2009ii,Wang:2009rw,cosmology38,cosmology39}.
The initial singularity is avoided because of
``dark" radiation, which is a negative $a^{-4}$ term.
It comes from higher curvature terms. 
The ``dark" radiation was first introduced 
in the context of a brane world~\cite{DR_BW}.
Although such an effect is very interesting and important,
the ``dark" radiation term may fail to avoid a singularity
if one include radiation or massless field.
The conventional radiation behaves as $a^{-4}$ with positive 
coefficient.
If we have a sufficient amount of real radiation,
the universe will inevitably collapse to
 a big-crunch singularity.
Furthermore, if we assume that radiation field has also the same scaling law 
as gravity in the UV limit, the energy density of radiation field 
changes as $a^{-6} $\cite{cosmology3},
 which is the same scaling law of stiff matter
in the conventional theory.
Inclusion of the positive $a^{-6}$ term will 
kill the possibility of singularity avoidance
by ``dark" radiation.
In order to save the present mechanism for singularity avoidance,
one needs a negative $a^{-6}$ term,
which may be obtained in the 
generalized HL gravity model \cite{SVW}. 

Recently some papers have discussed the case without the detailed balance
and studied a singularity avoidance 
(a bounce universe or an oscillating behavior):
One is by use of a phase space analysis
\cite{Carloni:2009jc,CaiSar},
and the other is the case with perfect fluid with time-evolving
equation of state~\cite{cosmology33,emergent_universe}.
The former analysis was not properly performed because they introduce
the dynamical variables more than the degrees of freedom.
In the latter case, although they discuss some interesting 
transitions, the assumption of the equation of state is not so clear.

In the present paper, since there has so far not been a systematic and 
substantial analysis in 
 cosmology based on this most general potential
without the detailed balance condition,
we provide a complete classification of the cosmological dynamics.
We do not include any matter fields not only 
for simplicity but also to avoid unclear assumption.
It is just straightforward to include 
perfect fluid with the equation of state $P=w\rho$ ($w$=constant).
In particular, our analysis includes matter fluid with radiation and
stiff matter as it is.
We clarify which conditions should be satisfied 
for singularity avoidance.
We also propose some possible scenario for a cyclic universe, i.e.,
the oscillating spacetime will transit by quantum tunneling 
to an inflationary phase, resulting in a cyclic universe after reheating.


The  paper is organized as follows.
After giving the generalized model of  Ho\v{r}ava-Lifshitz gravity 
in \S.\ref{Horava-Lifshitz},
we study the isotropic and homogeneous vacuum spacetime
in \S.\ref{FLRW}.
We find a variety of phase structures including 
a bounce universe and an oscillating universe.
We then invoke a more realistic cosmological model which may lead to
a macroscopic cyclic universe via quantum tunneling from an oscillating 
universe.

\section{Ho\v{r}ava-Lifshitz gravity and the coupling constants}
\label{Horava-Lifshitz}
The basic variables in HL gravity are
the lapse function, $N$, the shift vector, $N_i$,
and the spatial metric, $g_{ij}$.
These variables are subject to the action~\cite{Horava, SVW}
\begin{eqnarray}
S={1\over 2\kappa^2}
\int dt d^3x  \sqrt{g}N\left(
\mathscr{L}_K-\mathscr{V}_{\rm HL}[g_{ij}]
\right),
\end{eqnarray}
where $\kappa^2=1/M_{\rm PL}^2$ and
the kinetic term is given by
\begin{eqnarray}
\mathscr{L}_K=\mathpzc{K}_{~ij}\mathpzc{K}^{\,ij}-\lambda \mathpzc{K}^2
\end{eqnarray}
with 
\begin{eqnarray}
\mathpzc{K}_{~ij}:={1\over 2N}
\left(\dot g_{ij}-\nabla_iN_j-\nabla_jN_i\right)
\end{eqnarray}
being the extrinsic curvature.
The potential term $\mathscr{V}_{\rm HL}$ will be defined shortly.
In general relativity we have $\lambda=1$, only for which
the kinetic term is invariant under general coordinate transformations.
In HL gravity, however, Lorentz symmetry is broken in exchange for 
renormalizability
and the symmetry of the theory is invariance under
the foliation-preserving diffeomorphism transformations,
\begin{eqnarray}
t\to\bar t(t),\,~~ x^i\to\bar x^i(t,x^j).\label{f-p}
\end{eqnarray}

As implied by the symmetry ($\ref{f-p}$)
it is most natural to consider the projectable version of HL gravity,
for which the lapse function is dependent only on $t$: $N=N(t)$~\cite{Horava}.
Since the Hamiltonian constraint is derived from the variation with respect to
 the lapse function,
in the projectable version of the theory
the resultant constraint equation is not imposed locally at each point 
in space,
but rather is an integration over the whole space.
In the cosmological setting, the projectability condition
results in an additional dust-like component in the Friedmann equation 
[see Eq.~(\ref{Friedmann_eq}) below]~\cite{DM}.

The most general form of the potential $\mathscr{V}_{\rm HL}$ is given 
by~\cite{SVW}
\begin{eqnarray}
\mathscr{V}_{\rm HL}&=&
2\Lambda
+g_1\mathpzc{R}
\nonumber\\&&
+\kappa^2\left(g_2\mathpzc{R}^2
+g_3\mathpzc{R}^{i}_{~j}\mathpzc{R}^{j}_{~i}\right)
+\kappa^3 g_4\epsilon^{ijk}\mathpzc{R}_{\,i\ell}\nabla_j\mathpzc{R}^\ell_{~k}
\nonumber \\
&&+\kappa^4\Bigl(g_5\mathpzc{R}^3
+g_6\mathpzc{R}\,\mathpzc{R}^{i}_{~j}\mathpzc{R}^{j}_{~i}
+g_7\mathpzc{R}^{i}_{~j}\mathpzc{R}^{j}_{~k}\mathpzc{R}^{k}_{~i}
\nonumber\\&&
+g_8 \mathpzc{R}\Delta \mathpzc{R}
+g_9\nabla_i\mathpzc{R}_{\,jk}\nabla^i\mathpzc{R}^{jk}
\Bigr)
\,,\label{potential}
\end{eqnarray}
where $\Lambda$ is a cosmological constant,
$\mathpzc{R}^{i}_{~j}$
 and $\mathpzc{R}$ are the Ricci and scalar curvatures
of the 3-metric $g_{ij}$, respectively,
and
$g_i$'s ($i=1,..., 9$) are the dimensionless 
coupling constants.
(See Appendix A for some conditions on these coupling constants.)


In the original proposal~\cite{Horava} Ho\v{r}ava assumed the
detailed balance condition, by which the potential term~(\ref{potential})
is simplified to some extent.
The potential under the detailed balance condition is given by
\begin{eqnarray}
\mathscr{V}_{\rm DB}&=&
-{3\kappa^2\mu^2\Lambda_{W}^2\over 2(3\lambda-1)}
+{\kappa^2\mu^2\Lambda_W\over 2(3\lambda-1)}\mathpzc{R}
\nonumber\\&&
-{(4\lambda-1)\kappa^2\mu^2
\over 8(3\lambda-1)}\mathpzc{R}^2
+{\kappa^2\mu^2\over 2}\mathpzc{R}_{~i}^{j}\mathpzc{R}_{~j}^{i}
\nonumber\\&&
- {2\kappa^2\mu\over \omega^2}
\mathpzc{C}^{\,ij}\mathpzc{R}_{~ij}
+{2\kappa^2\over \omega^4}\mathpzc{C}_{\,ij}\mathpzc{C}^{\,ij}
\,,\label{db-pot}
\end{eqnarray}
where 
\begin{eqnarray}
\mathpzc{C}^{\,ij}:=\epsilon^{ik\ell}\nabla_k 
\left(\mathpzc{R}^j_{~\ell}-{1\over 4}
 \mathpzc{R} \delta^j_{~\ell}\right)
\end{eqnarray}
is
the Cotton tensor, and 
$\Lambda_W$, $\mu$ and $\omega$ are constants.
The potential~(\ref{db-pot}) is therefore reproduced by identifying
\begin{eqnarray}
&&\Lambda=
-{3(3\lambda-1)\over 2\mu^2\kappa^2},
\label{cosmological_constant}
\\
&&g_1 = -1,
\label{R1}
\\
&&g_2=-{(4\lambda-1)
\over 4(3\lambda-1)}\mu^2\kappa^2,\quad
g_3\,=\,\mu^2\kappa^2\,,
\label{R2}
\\
&&g_4 = - {4\mu\kappa^2\over\omega^2},\quad
g_5\,=\,{2\kappa^2\over \omega^4},\quad
g_6\,=\,-{10\kappa^2\over \omega^4}\,,
\nonumber \\
&&g_7={12\kappa^2\over \omega^4},\quad
g_8\,=\,{3\kappa^2\over 2 \omega^4},\quad
g_9\,=\,{4\kappa^2\over  \omega^4}
\label{R3}
\,,
\end{eqnarray}
and $\Lambda_W=-(3\lambda-1)/(\mu^2\kappa^2)$.
In the detailed balance case $\mu$ and $\omega$ are two free parameters.

In what follows, we adopt the unit of $\kappa^2=1 (M_{\rm PL}^2=1)$
for brevity.

\section{FLRW universe in Ho\v{r}ava-Lifshitz gravity}
\label{FLRW}
We discuss an isotropic and homogeneous 
vacuum universe in Ho\v{r}ava-Lifshitz gravity.
Note that such a vacuum spacetime
 is not realized in general relativity.
We will extend our analysis to anisotropic spacetime
(Bianchi cosmology) in the separate paper.

Assuming a 
FLRW spacetime,
 which metric is given by
\begin{eqnarray}
ds^2=-dt^2+a^2 \left({dr^2\over 1-Kr^2}+r^2d\Omega^2\right)
\,,
\end{eqnarray}
with $K=0$ or $\pm 1$.
We find the Friedmann equation as
\begin{widetext}
\begin{eqnarray}
H^2+{2\over (3\lambda-1)}{K\over a^2}
={2\over 3(3\lambda-1)}\left[\Lambda+{g_{\rm d}\over a^3}
+{g_{\rm r}\over a^4}
+{g_{\rm s}\over a^6}
\right]
\label{Friedmann_eq}
\,,
\end{eqnarray}
\end{widetext}
where $H=\dot{a}/a$, 
\begin{eqnarray}
g_{\rm d}&:=&8C\,.
\nonumber \\
g_{\rm r}&:=&6(g_3+3g_2)K^2\,,
\nonumber \\
g_{\rm s}&:=&12(9g_5+3g_6+g_7)K^3\,.
\end{eqnarray}
A constant $C$ may appear from the projectability 
condition and
could be ``dark matter"\cite{DM}.
For a flat universe ($K=0$), the higher curvature terms do not give
any contribution, and then the dynamics is almost trivial.
Hence, in this paper, we discuss only non-flat universe ($K=\pm 1$).

If $\lambda=1$, we find a usual 
Friedmann equation for an isotropic and homogeneous universe
in GR with a cosmological constant, 
dust, radiation and stiff matter.
If $g_{\rm d}, g_{\rm r}$, and $g_{\rm s}$ are non-negative,
such a spacetime gives a conventional FLRW universe model.
However, since  those coefficients come from higher curvature terms,
their positivity is not guaranteed. Rather some of them could be negative.
As a result, we find an unconventional cosmological scenario,
which we shall discuss here.
In what follows, we assume that $\lambda>1/3$, but
do not fix it to be unity.

In this paper, we 
assume $C=0$ just for simplicity.
The Friedmann equation is written as
\begin{eqnarray}
{1\over 2}\dot{a}^2+\mathpzc{U}(a)=0
\label{Friedmann_eq2}
\,,
\end{eqnarray}
where 
\begin{eqnarray}
\mathpzc{U}(a)={1\over 3\lambda-1}\left[
K-{\Lambda\over 3}a^2
-{g_{\rm r}\over 3a^2}-{g_{\rm s}
\over 3a^4}\right]
\,.
\end{eqnarray}

Since the scale factor $a$ changes as
a particle with zero energy in this ``potential" $\mathpzc{U}$,
the condition $\mathpzc{U}(a)\leq 0$ gives the possible range 
of $a$ when the universe evolves. So 
we can classify the ``motion" of the universe by the signs of 
 $K$ and $\Lambda$, and by the values of $g_{\rm r}$ and $g_{\rm s}$.
Note that in the case with the detailed balance condition,
we have
\begin{eqnarray}
&& g_{\rm r}=6(g_3+3g_2)=-{3\mu^2\over 2(3\lambda-1)}<0 ~~{\rm for}
~~~\lambda>1/3
\nonumber\\
&& g_{\rm s}=12(9g_5+3g_6+g_7)K=0
\label{grgs_DB}
\,.
\end{eqnarray}
It is some special case of our analysis, although
its dynamics will be completely different from generic cases
because $g_{\rm s}$ vanishes.

We find mainly the following four types of the FLRW universe:\\[1em]
(1) [$\mathpzc{BB}$
$\Rightarrow$$\mathpzc{BC}$]: Suppose $\mathpzc{U}(a)\leq 0$ for 
$a\in (0,a_T]$,
and the equality is true only when $a=a_T$.
A spacetime starts from a big bang ($\mathpzc{BB}$) and expands, but
it eventually turns around at $a=a_T$ to contract, finding a big crunch ($\mathpzc{BC}$).
$a_T$ is a scale factor when the universe turns around
from expansion to contraction.
\\[.5em]
(2)[$\mathpzc{BB}\Rightarrow \infty$ or $\infty\Rightarrow\mathpzc{BC}$]:
If $\mathpzc{U}(a)< 0$ for any positive values of $a$,
a spacetime starts from a big bang and expands forever, 
or its time reversal (A spacetime contracts 
to a big crunch). As for the asymptotic spacetime, we find 
 $a\propto t$ [Milne:$\mathpzc{M}$] ($K=-1$) for $\Lambda=0$,
 while $a\propto \exp(\sqrt{\Lambda/3}\, t)$ [de Sitter:$\mathpzc{dS}$]
 for $\Lambda>0$.
We denote them as   
$\mathpzc{BB}\Rightarrow \mathpzc{M}$,
 and $\mathpzc{BB}\Rightarrow \mathpzc{dS}$, respectively.
For the contracting cases, we describe them as 
$\mathpzc{M}\Rightarrow \mathpzc{BC}$, and $\mathpzc{dS}\Rightarrow
 \mathpzc{BC}$, respectively.\\[.5em]
(3) [$\mathpzc{Bounce}$]:
If $\mathpzc{U}(a)\leq 0$ for 
$a\in [a_T,\infty)$
and the equality holds only when $a=a_T$,
a spacetime initially contracts from an infinite scale,
and it eventually turns around at a finite scale $a_T$,
and expands forever. 
The asymptotic spacetimes are the same as the case (2):
$\mathpzc{M}$, and $\mathpzc{dS}$.\\[.5em]
(4) [$\mathpzc{Oscillation}$]:
If $\mathpzc{U}(a)\leq 0$  for 
$a\in [a_{\rm  min},a_{\rm  max}]$
and the equality holds only when $a=a_{\rm min}$ and $a=a_{\rm max}$,
a spacetime oscillates between two finite scale factors.

For some specific values (or specific relations)
of $g_{\rm r}$ and $g_{\rm s}$, which divides
two different phases of spacetimes,
we find a static universe ($\mathpzc{S}$):\\
(5) [$\mathpzc{S}$]: A spacetime is static
 with a constant scale factor $a_S$, 
if $\mathpzc{U}(a_S)=0$ and $\mathpzc{U}'(a_S)=0$.

There are two types of static universes: one is stable 
($\mathpzc{S}_{\rm s}$)
and the other is
unstable ($\mathpzc{S}_{\rm u}$).
When we have an unstable static universe, we also find the following
 types of dynamical universes with a static spacetime as
an asymptotic state as well:\\
(6) [$\mathpzc{S}_{\rm u}$ $\Rightarrow \infty$ or 
$\infty\Rightarrow$$\mathpzc{S}_{\rm u}$]:
If $\mathpzc{U}(a)\leq 0$ for 
$a\in [a_S,\infty)$
and the equality holds only at $a_S$,
a spacetime 
starts from a static state in the infinite past, and expands forever, or it
initially contracts from an infinite scale, and eventually 
reach  a static state in the infinite future.
We then have $\mathpzc{S}_{\rm u}$
 $\Rightarrow \mathpzc{dS}$, $\mathpzc{M}$ or 
$\mathpzc{dS}$, $\mathpzc{M}\Rightarrow$$\mathpzc{S}_{\rm u}$\\[.5em]
(7) [$\mathpzc{BB}$$\Rightarrow$ $\mathpzc{S}_{\rm u}$,
 or $\mathpzc{S}_{\rm u}$ $\Rightarrow$ $\mathpzc{BC}$]:
If 
$\mathpzc{U}(a)\leq 0$ for 
$a\in (0,a_S]$
and the equality holds only at $a_S$,
 A spacetime starts from a big bang and expands to
 a static state with a finite scale $a_S$, or 
its time reversal (A spacetime contracts from a static state
to a big crunch).\\[.5em]
(8) [$\mathpzc{S}_{\rm u}$ $\Rightarrow$ $\mathpzc{Bounce}$
 $\Rightarrow$ $\mathpzc{S}_{\rm u}$]: 
If 
$\mathpzc{U}(a)\leq 0$ for 
$a\in [a_S, a_T]$ (or $a\in [a_T,a_S]$) 
and the equality holds only at $a_S$ and
$a_T$,
a spacetime 
starts from a static state in the infinite past, and expands (or contracts).
It eventually bounces at a finite scale 
$a_T$, and then reach a static state again
in the infinite future.

For the case of $\Lambda\neq 0$,
introducing the curvature scale $\ell$
which is defined by
\begin{eqnarray}
{\Lambda\over 3}={\epsilon \over \ell^2}\,,
\end{eqnarray}
where $\epsilon=\pm 1$,
we can rescale the variables and 
rewrite
the ``potential" $\mathpzc{U}$ 
by the rescaled variables as
\begin{eqnarray}
\mathpzc{U}(a)={1\over 3\lambda-1}\left[
K-\epsilon \tilde a^2
-{\tilde g_{\rm r}\over 3\tilde a^2}-{\tilde g_{\rm s}
\over 3\tilde a^4}\right]
\,,
\end{eqnarray}
where
$\tilde a=a/\ell$, $\tilde  g_{\rm r}= g_{\rm r}/\ell^2$,
and $\tilde  g_{\rm s}= g_{\rm s}/\ell^4$.
Using this potential and variables,
we can discuss the fate of the universe
without specifying the value of $\Lambda$.

A static universe will appear if we find 
a solution $a=a_S(>0)$
which satisfies  $\mathpzc{U}(a_S)=0$ and $\mathpzc{U}'(a_S)=0$.
If $\Lambda\neq 0$ ($\epsilon=\pm 1$),
it happens if there is
 a relation between $\tilde g_{\rm r}$ and $\tilde g_{\rm s}$,
which is defined by 
\begin{eqnarray}
\tilde g_{\rm s}
&\,=&
\tilde g_{\rm s}^{\,[\epsilon, K](\pm)}\left(\tilde g_{\rm r}\right)
\nonumber \\
&:=&
{1\over 9\epsilon^2}\left[
2K-3\epsilon K\tilde g_{\rm r}\pm 2 (1-\epsilon \tilde g_{\rm r})^{3/2}\right]
\,.~~~
\label{grgs_L}
\end{eqnarray}
This gives the curve $\Gamma_{\epsilon, K(\pm)}$ 
on the $\tilde g_{\rm r}$-$\tilde g_{\rm s}$ plane, which 
gives the boundary between two different phases of spacetime.
The radius of a static universe is given by
\begin{eqnarray}
\tilde a_S
&=&\tilde a_S^{[\epsilon,K](\pm)}:=
\sqrt{
{1\over 3\epsilon}\left[
K\pm\sqrt{1-\epsilon \tilde g_{\rm r}}\right]
}
\label{aS_L}
\,,
\end{eqnarray}
if it is real and positive.
Here $\pm$ correspond to the curves $\Gamma_{\epsilon, (\pm)}$.
Since $\mathpzc{U}(\tilde a)=0$ is the cubic equation
with respect to $\tilde a^2$ and $\tilde a^2=\tilde a_S^2$ is the double root,
we have the third root, which is given by
\begin{eqnarray}
\tilde a_T
&=&\tilde a_T^{[\epsilon,K](\pm)}:=
\sqrt{
{1\over 3\epsilon}\left[
K\mp  2 \sqrt{1-\epsilon \tilde g_{\rm r}}\right]
}
\label{aT_L}
\,,
\end{eqnarray}
where the universe turns around (or bounces).
To exist such a point, it must be real and positive.

If $\Lambda=0$, we find 
\begin{eqnarray}
g_{\rm s}
=
-{K\over 12} g_{\rm r}^2
\,,
\label{grgs_L0}
\end{eqnarray}
which is found from (\ref{grgs_L})
in the limit of $\epsilon =0$.
The corresponding curve on the $g_{\rm r}$-$g_{\rm s}$ plane
is denoted by $\Gamma_{0,K}$.
The radius is given by
\begin{eqnarray}
a_S
&=&a_S^{[0,K]}:=
\sqrt{
{g_{\rm r}\over 6K}
}
\label{aS_L0}
\,,
\end{eqnarray}
assuming $Kg_{\rm r}>0$.

Note that our classification depends just on 
$g_{\rm r}$ and $g_{\rm s}$ (or $\tilde  g_{\rm r}$ and $\tilde  g_{\rm s}$),
apart from $K$ and $\Lambda$.
Since $g_{\rm r}$ and $g_{\rm s}$  are given by $g_2, g_3, g_5, g_6$ 
and $g_7$, but do not include $g_4, g_8$ and $g_9$,
the fate of the universe is classified only by
the conditions on the coupling constants
of higher curvature terms but not on those of their derivatives
such as $\nabla_j\mathpzc{R}^\ell_{~k}$.

In the case with the detailed balance conditions,
we find $\Lambda<0$ from Eq. (\ref{cosmological_constant}),
and then obtain from Eq. (\ref{grgs_DB}) 
\begin{eqnarray}
\tilde  g_{\rm r}=-9/4
\,,~~~
\tilde  g_{\rm s}=0
\,.
\end{eqnarray}

Now we shall discuss what kind of spacetimes are
realized under which conditions 
in the following three cases  separately 
[{\bf A.} $\Lambda=0$, {\bf B.}  $\Lambda>0$,
and {\bf C.} $\Lambda<0$].

\subsection{$\Lambda=0$}
\label{Lambda0}
\begin{figure}[h]
\begin{center}
\includegraphics[scale=.4]{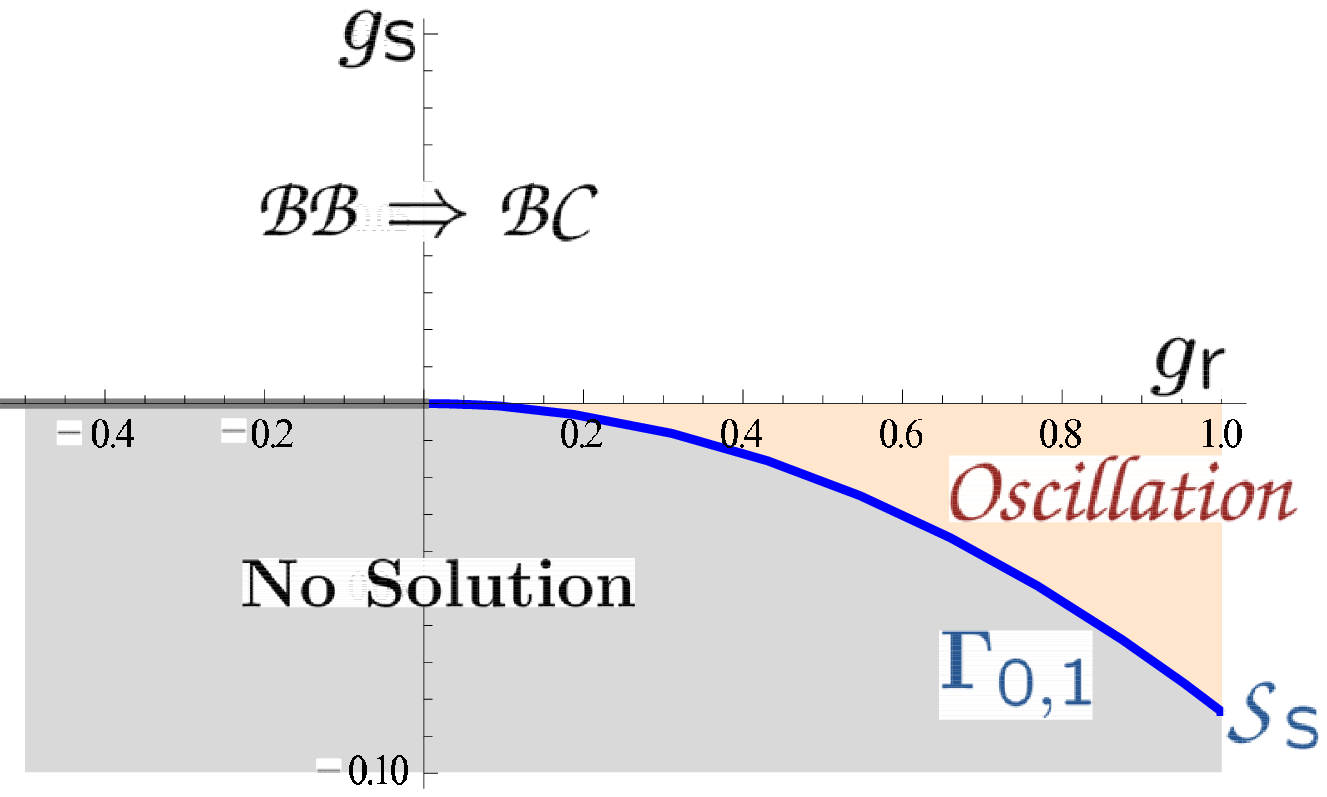}
\\
(a) $K=1$
\\[2em]
\includegraphics[scale=.4]{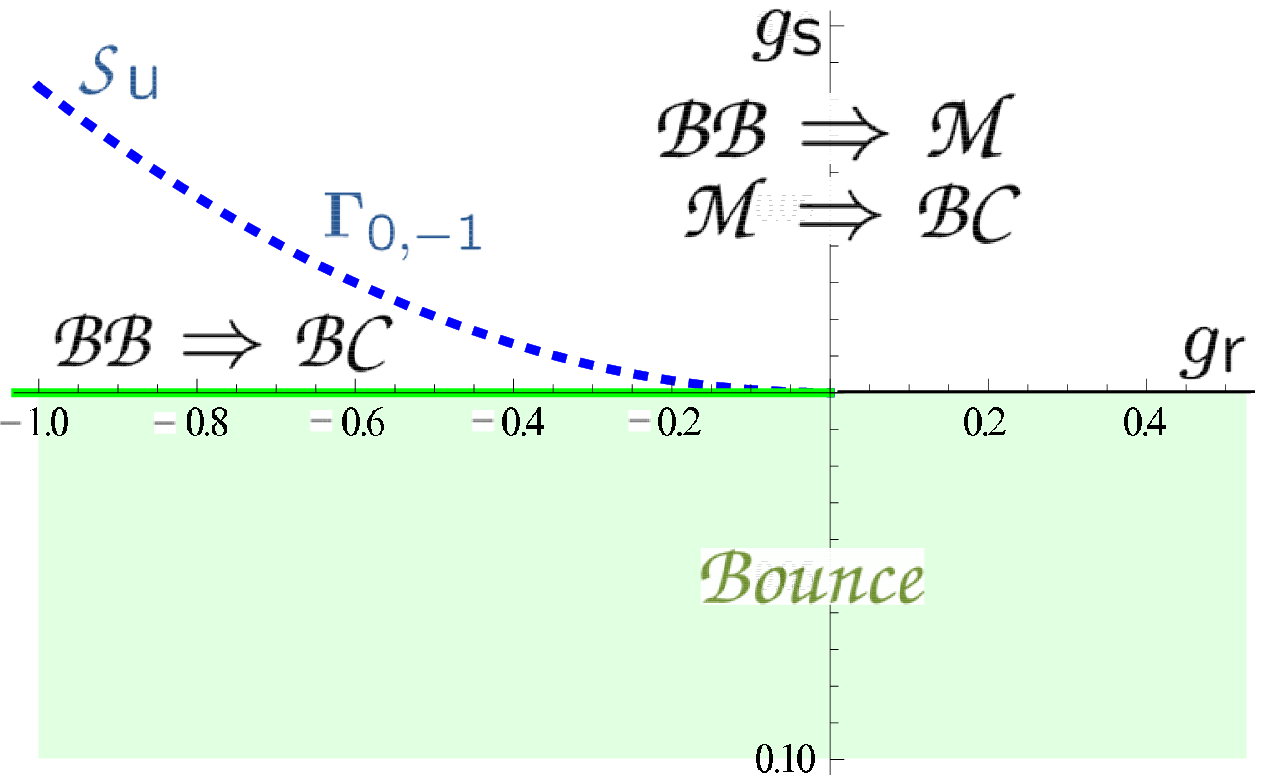}
\\
(b) $K=-1$
\caption{Phase diagram of spacetimes for $\Lambda=0$.
The oscillating universe is found only for the case of $K=1$.
The stable and unstable 
static universes ($\mathpzc{S}_{\rm s}$ 
and $\mathpzc{S}_{\rm u}$) exist on the boundary $\Gamma_{0,1}$ and
$\Gamma_{0,-1}$, respectively. 
On $\Gamma_{0,-1}$, we also find dynamical
universes with an asymptotically static spacetime;
$\mathpzc{S}_{\rm u}$ $\Rightarrow$ $\mathpzc{BC}$,
$\mathpzc{S}_{\rm u}$ $\Rightarrow$ $\mathpzc{M}$,
$\mathpzc{BB}$$\Rightarrow$ $\mathpzc{S}_{\rm u}$,
 or $\mathpzc{M}$$\Rightarrow$ $\mathpzc{S}_{\rm u}$.
}
\label{fig:Lambda0}
\end{center}
\end{figure}

If a cosmological constant is absent, the ``potential" is written as
\begin{eqnarray}
\mathpzc{U}(a)
&=&
{1\over (3\lambda-1)a^4}\left[
Ka^4-{g_{\rm r}\over 3}a^2-{g_{\rm s}\over 3}
\right]
\,.
\end{eqnarray}
In Fig. \ref{fig:Lambda0}, we show the fate of the universe, which depends on
the values of $ g_{\rm r}$ and $g_{\rm s}$.
For the case of $K=1$,
there are two types of spacetime  phases: One is
$\mathpzc{BB}\Rightarrow\mathpzc{BC}$, 
and the other is an oscillating universe.
In fact, 
if $g_{\rm r}>0, g_{\rm s}<0$ and $g_{\rm r}^2+12g_{\rm s}>0$,
we find the scale factor 
$a$ is bounded in a finite range 
as $(0<) ~a_{\rm min}\leq a\leq a_{\rm max} ~(<\infty)$, 
where
\begin{eqnarray}
a_{\rm min}^2&\equiv &{1\over 6}\left[
g_{\rm r}-\sqrt{g_{\rm r}^2+12g_{\rm s}}
\right]
\nonumber \\
a_{\rm max}^2&\equiv &{1\over 6}\left[
g_{\rm r}+\sqrt{g_{\rm r}^2+12g_{\rm s}}
\right]
\label{aminmax}
\,,
\end{eqnarray}
which gives an oscillating universe.
The condition for an  oscillating universe is written as
\begin{eqnarray}
g_{\rm r}>0~\,,~~~
-{g_{\rm r}^2\over 12}\leq g_{\rm s}<0
\,,
\label{cond_0}
\end{eqnarray}
which is shown in Fig. \ref{fig:Lambda0}(a)
 by ``$\mathpzc{Oscillation}$" (the light-orange colored region)  
in the $g_{\rm s}$-$g_{\rm r}$ plane.
The equality in Eq. (\ref{cond_0}), which is
the curve $ \Gamma_{0,1}$,
gives a static universe with the scale factor 
$a=a_S:=\sqrt{g_{\rm r}/6}$.

For the case of $K=-1$, we find three types of spacetime phases:
$\mathpzc{BB}\Rightarrow \mathpzc{M}$ 
(or $\mathpzc{M}\Rightarrow \mathpzc{BC}$),
$\mathpzc{BB}\Rightarrow \mathpzc{BC}$, 
 and $\mathpzc{Bounce}$ (see Fig.\ref{fig:Lambda0}(b)).
On the boundary curve $\Gamma_{0,-1}$,
which is defined by 
Eq. (\ref{grgs_L0}), i.e.,
$g_{\rm s}=g_{\rm r}^2/12$ ($g_{\rm r}<0$),
we find an unstable static universe $\mathpzc{S}_{\rm u}$, and 
the dynamical
universes with an asymptotically static spacetime;
$\mathpzc{S}_{\rm u}$ $\Rightarrow$ $\mathpzc{BC}$,
$\mathpzc{S}_{\rm u}$ $\Rightarrow$ $\mathpzc{M}$,
$\mathpzc{BB}$$\Rightarrow$ $\mathpzc{S}_{\rm u}$,
 or $\mathpzc{M}$$\Rightarrow$ $\mathpzc{S}_{\rm u}$.

The bounce universe is found if $g_{\rm s}<0$ or 
$g_{\rm s}=0$ with $g_{\rm r}<0$, which is shown
by "$\mathpzc{Bounce}$" (the light-green region).
The radius at a turning point, $a_{T}$ is given by
\begin{eqnarray}
a_{T}
=\sqrt{
{1\over 6}\left(-g_{\rm r}+\sqrt{g_{\rm r}^2-12
g_{\rm s}}\right)
}
\label{aT_L0}
\end{eqnarray}

Next we shall evaluate the period of
an oscillating universe in the case of $K=1$.
The solution for Eq. (\ref{Friedmann_eq2}) is given by
\begin{eqnarray}
t-t_{\rm max}&=&-\int_{a_{\rm max}}^{a}{da\over 
\sqrt{-2\mathpzc{U}(a)}}
\nonumber \\
&=&
a_{\rm max}\,\sqrt{3\lambda-1\over 2}\,
E\left(\phi[a],k\right)
\,,~~~
\end{eqnarray}
where 
$E(\phi, k)$ is the elliptic integral of the second kind,
which is defined by
\begin{eqnarray}
E(\phi, k):=\int_0^{\phi} d\theta\, \sqrt{1-k^2\sin^2 \theta}
\,.
\end{eqnarray}
$k$ and $\phi[a]$ are given by
\begin{eqnarray}
k:&=&{\sqrt{a_{\rm max}^2-a_{\rm min}^2}\over
 a_{\rm max}}
\,,
\\
\phi[a]:&=&\sin^{-1}\left({a_{\rm max}^2-a^2\over
a_{\rm max}^2-a_{\rm min}^2}\right)
\,.
\end{eqnarray}
The period $T$ is given by
\begin{eqnarray}
T:=2(t_{\rm min}-t_{\rm max})=
2{a_{\rm max}}\,\sqrt{3\lambda-1\over 2}\,
E\left(k\right)
\,,~~~
\end{eqnarray}
where $E(k)$ is the complete elliptic integral of the second kind
 defined by $E(k):=E(\pi/2,k)$.

\begin{figure}[h]
\begin{center}
\includegraphics[scale=.4]{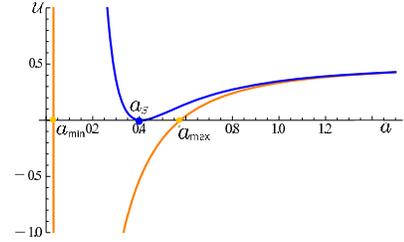}
\caption{The potential 
$\mathpzc{U}(\tilde a)$
 for a stable static universe and an oscillating universe
 near the $g_{\rm r}$-axis. The ``coupling" constants are chosen as
$g_{\rm r}= 1$ and $g_{\rm s}=-1/12$ on $\Gamma_{0,1}$ for a static
universe, which radius is shown by $a_S=1/\sqrt{6}$.
We also show the case with  
$g_{\rm r}= 1$ and $g_{\rm s}=-0.001$
for an oscillating universe, which maximum and minimum radii
are given by $a_{\rm max}=0.576481$ 
and $a_{\rm min}=0.0316705$, respectively.
}
\label{fig:pot_L0kp}
\end{center}
\end{figure}

In order to evaluate the period, we consider some limiting cases,
which are the boundaries of the region of $\mathpzc{Oscillation}$.
In Fig. \ref{fig:pot_L0kp},
we show the potential $\mathpzc{U}(a)$ by the blue curve
 for one boundary curve $\Gamma_{0,1}$,
which is given by 
$g_{\rm s} =-g_{\rm r}^2/12$.
It gives a stable static universe 
with the scale factor $a_S$.
We also show the potential 
near the other boundary of $\mathpzc{Oscillation}$
(the positive $g_{\rm r}$-axis)
by the dashed orange curve.
Choosing, for example, $g_{\rm r}= 1$ and $g_{\rm s}=-0.001$, 
we find an oscillating univese with the scale factor 
$a\in [0.0316705, 0.576481]$.
Since these two potentials give the limiting cases,
we find that $0<a_{\rm min}\leq a_S$ and $a_S \leq a_{\rm max} <
\sqrt{2}\, a_S$
for an oscillating universe. 

In the limit of a static universe (near $\Gamma_{0,1}$), we find
the period $T_S$ as
\begin{eqnarray}
T_S=\pi
\sqrt{\left({3\lambda-1\over 2}\right){g_{\rm r}\over 6}}
\,,
\end{eqnarray}
while in the other boundary limit ($g_{\rm s}\rightarrow 0$),
we obtain 
\begin{eqnarray}
T_0=
\sqrt{\left({3\lambda-1\over 2}\right){4g_{\rm r}\over 3}}
\,.
\end{eqnarray}
From these evaluations, 
giving the value of $g_{\rm r}$, we find the period $T$ 
of any oscillating universe is 
bounded in the range of 
$(T_0, T_S)$ for $g_{\rm s}\in (-g_{\rm r}^2/12,0)$.
We then approximate the period as $T\sim g_{\rm r}^{1/2}$.

We have found an oscillating FLRW universe because 
we have ``negative" energy of ``stiff matter" which comes 
from the higher curvature term. 
The condition for an  oscillating universe is rewritten 
in terms of the original coupling constants as
\begin{eqnarray}
&&
g_3+3g_2>0\,,
\\
&&
-{(g_3+3g_2)^2\over 4}\leq 9g_5+3g_6+g_7<0
\,.
\end{eqnarray}

\subsection{$\Lambda>0 ~(\epsilon=1)$}
\label{Lambda+}
\begin{figure}[h]
\begin{center}
\includegraphics[scale=.4]{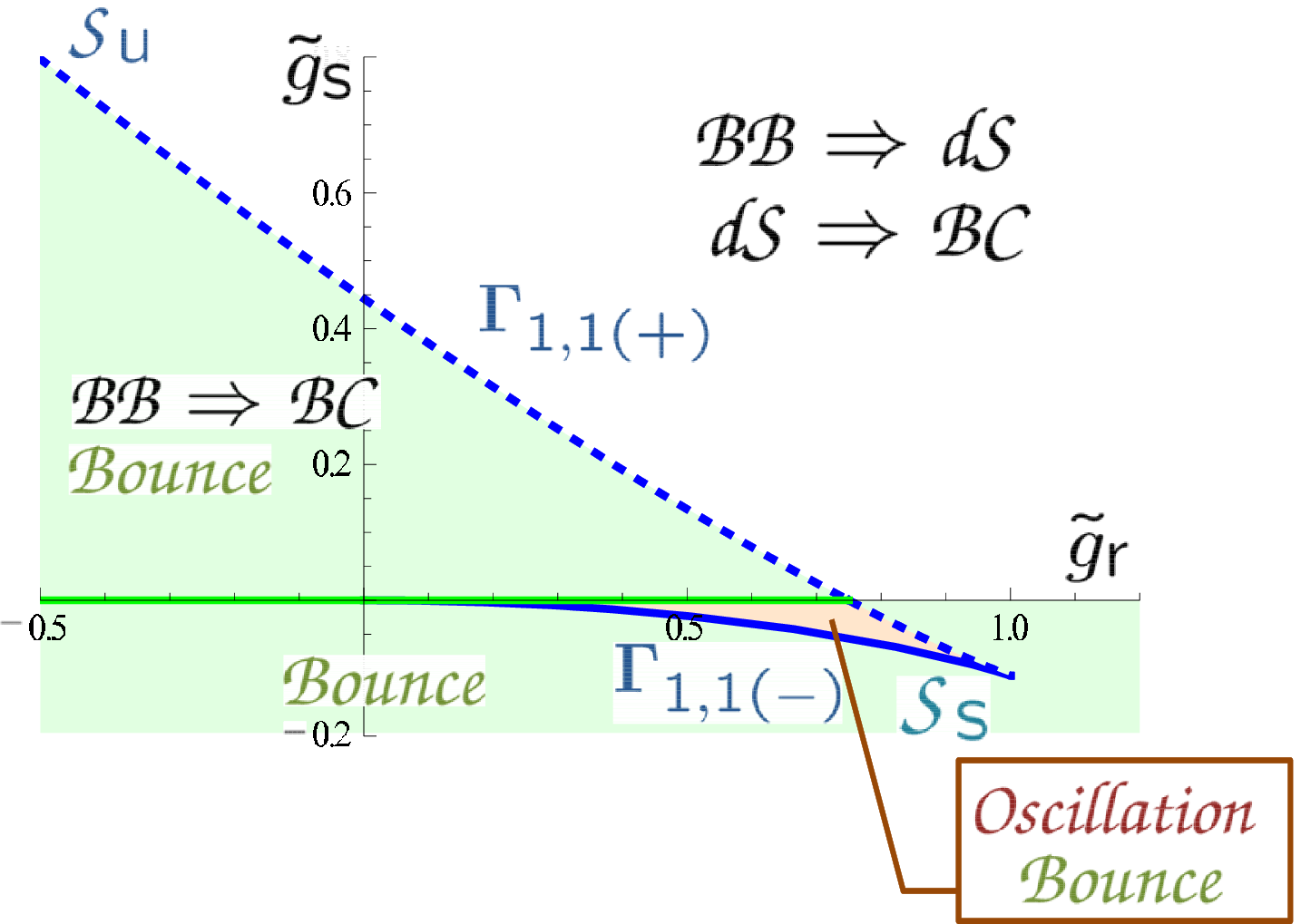}
\\
(a) $K=1$
\\[2em]
\includegraphics[scale=.4]{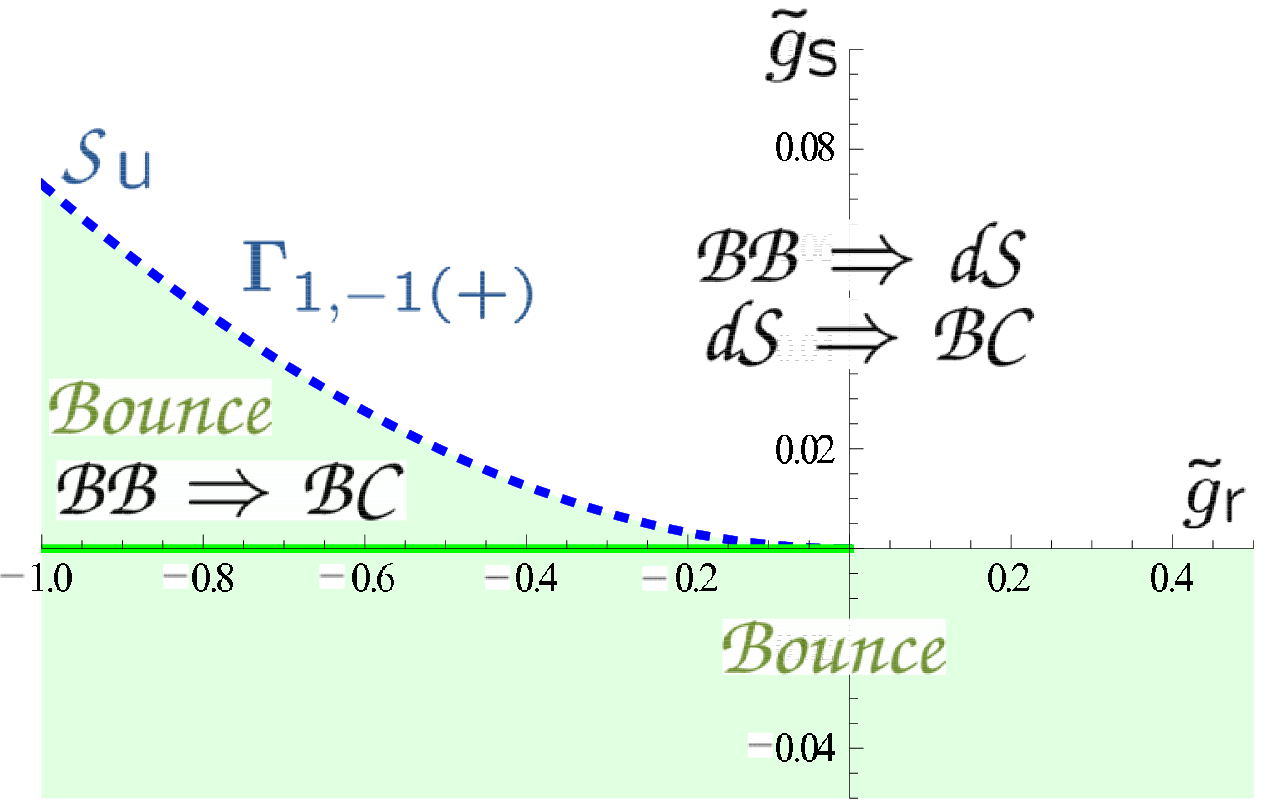}
\\
(b) $K=-1$
\caption{Phase diagram of spacetimes for $\Lambda>0$.
The oscillating universe is found only for the case of $K=1$.
The static universes ($\mathpzc{S}_{\rm s}$ and $\mathpzc{S}_{\rm u}$) 
exist on the boundaries $\Gamma_{1,K(\pm)}$.
 We also find dynamical
universes with an asymptotically static spacetime;
$\mathpzc{S}_{\rm u}$ $\Rightarrow$ $\mathpzc{dS}$ or 
$\mathpzc{S}_{\rm u}$ $\Rightarrow$ $\mathpzc{BC}$ 
on $\Gamma_{1,1(+)}(\tilde g_{\rm s}\geq 0)$;
$\mathpzc{S}_{\rm u}$ $\Rightarrow$ $\mathpzc{dS}$ or 
$\mathpzc{S}_{\rm u}$ $\Rightarrow$ $\mathpzc{Bounce}$
$\Rightarrow$ $\mathpzc{S}_{\rm u}$ 
on $\Gamma_{1,1(+)}(\tilde g_{\rm s}< 0)$;
$\mathpzc{S}_{\rm u}$ $\Rightarrow$ $\mathpzc{dS}$ or 
$\mathpzc{S}_{\rm u}$ $\Rightarrow$ $\mathpzc{BC}$ 
on $\Gamma_{1,-1}$.
}
\label{fig:Lambdap}
\end{center}
\end{figure}

In this case, the potential is given by
\begin{eqnarray}
\mathpzc{U}(\tilde a)={1\over (3\lambda-1)\tilde a^4}\left[
K\tilde a^4- \tilde a^6
-{\tilde g_{\rm r}\over 3}\tilde a^2-{\tilde g_{\rm s}
\over 3}\right]
\,.
\end{eqnarray}
For each value of $K$, 
we depict the fate of the universe
in Fig \ref{fig:Lambdap},
which depends on the values of $\tilde  g_{\rm r}$
and $\tilde  g_{\rm s}$.

We find non-singular evolution of the universe
($\mathpzc{Bounce}$, $\mathpzc{Oscillation}$, or 
$\mathpzc{Static}$) as well as 
the universe with a cosmological singularity
($\mathpzc{BB}$ $\Rightarrow$ $\mathpzc{BC}$,
$\mathpzc{BB}$ $\Rightarrow$ $\mathpzc{dS}$, or 
$\mathpzc{dS}$ $\Rightarrow$ $\mathpzc{BC}$).
Except for the case of $\mathpzc{BB}\Rightarrow\mathpzc{BC}$
and a static universe,
the expanding universe approaches de Sitter spacetime 
(exponentially expanding universe) because of a positive
cosmological constant $\Lambda$.  
The oscillating universe  exists if and only if 
$K=1$ and the following conditions are satisfied:
\begin{eqnarray}
&&
\tilde g_{\rm r}>0
\\
&&
\tilde g_{\rm s}^{\,[1,1]\rm (-)}(\tilde g_{\rm r})
\leq 
\tilde g_{\rm s}
\left\{
\begin{array}{cc}
<& 0\\[.5em]
\leq & \tilde g_{\rm s}^{\,[1,1]\rm (+)} (\tilde g_{\rm r})
\,,
\\
\end{array}
\right.
\end{eqnarray}
where $\tilde g_{\rm s}^{\,[1, 1] \rm (\pm)}$ is defined 
by Eq. (\ref{grgs_L}) with $\epsilon=1, K=1$.
This condition gives the constraint on $\tilde g_{\rm s}$ as
$-1/9\leq\tilde g_{\rm s}<0$.
Note that in the limit of $\tilde g_{\rm r}\ll 1$
(i.e. $\Lambda\rightarrow 0$), we recover
the condition (\ref{cond_0}).

The boundaries of two different phases of spacetimes consist of 
the $\tilde g_{\rm r}$-axis,  and  two curves 
($\Gamma_{1,1(\pm)}$)  for $K=1$
 or one curve ($\Gamma_{1,-1(+)}$) for $K=-1$.
Those boundary curves $\Gamma_{1,K(\pm)}$
are defined by
$\tilde g_{\rm s}=\tilde g_{\rm s}^{[1,K]\rm (\pm)}(\tilde g_{\rm r})$.

A stable static universe exist on the boundary curve  
$\Gamma_{1,1(-)}$, while 
 unstable static universes appear on the boundary curves
 $\Gamma_{1,\pm 1(+)}$.
For $K=1$, 
there are two types of static universes (stable and unstable)
corresponding to two curves 
$\Gamma_{1, 1(-)}$ and $\Gamma_{1, 1(+)}$, respectively,
which  coincide at $\tilde g_{\rm r}= 1$
 and $\tilde g_{\rm s}=-1/9$.
In the branches of unstable static universes
($\Gamma_{1,K(+)}$),
we also find dynamical
universes with an asymptotically static spacetime;
$\mathpzc{S}_{\rm u}$ $\Rightarrow$ $\mathpzc{dS}$ or 
$\mathpzc{S}_{\rm u}$ $\Rightarrow$ $\mathpzc{BC}$ 
on $\Gamma_{1,1(+)}(\tilde g_{\rm s}\geq 0)$;
$\mathpzc{S}_{\rm u}$ $\Rightarrow$ $\mathpzc{dS}$ or 
$\mathpzc{S}_{\rm u}$ $\Rightarrow$ $\mathpzc{Bounce}$
$\Rightarrow$ $\mathpzc{S}_{\rm u}$ 
on $\Gamma_{1,1(+)}(\tilde g_{\rm s}< 0)$;
$\mathpzc{S}_{\rm u}$ $\Rightarrow$ $\mathpzc{dS}$ or 
$\mathpzc{S}_{\rm u}$ $\Rightarrow$ $\mathpzc{BC}$ 
on $\Gamma_{1,-1(+)}$.

\begin{figure}[h]
\begin{center}
\includegraphics[scale=.4]{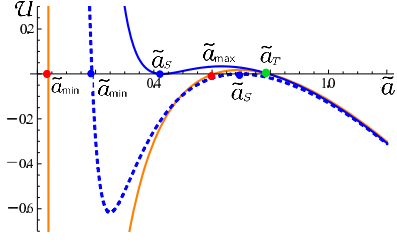}
\caption{The potential 
$\mathpzc{U}(\tilde a)$ for a stable and unstable static universes
 (the solid blue and the dashed blue),
 and that for an oscillating universe
 near $\tilde g_{\rm r}$-axis (orange). The constants are chosen as
$\tilde g_{\rm r}= 0.8$ and $\tilde g_{\rm s}=-0.0643206$
 on $\Gamma_{1,1(-)}$, and 
 $-0.0245683$ on $\Gamma_{1,1(+)}$, for static
universes, which radii are given by $\tilde a_S$, and 
$\tilde g_{\rm r}= 0.7$ and $\tilde g_{\rm s}=-0.001$
for an oscillating universe, which maximum and minimum radii
are given by $\tilde a_{\rm max}$ and $\tilde a_{\rm min}$, respectively.
We also find $\mathpzc{S}_{\rm u}$ $\Rightarrow$ $\mathpzc{Bounce}$
$\Rightarrow$ $\mathpzc{S}_{\rm u}$, which bounce radius is given by
$\tilde a_{\rm min}$. }
\label{fig:pot_Lpkp}
\end{center}
\end{figure}

The period $T$ of an oscillating universe is calculated by 
\begin{eqnarray}
\tilde T:&=&2\int_{\tilde a_{\rm min}}^{\tilde a_{\rm max}}
{d\tilde a\over \sqrt{-2\mathpzc{U}(\tilde a)}}
\,,
\label{period_nonzero_L}
\end{eqnarray}
where $\tilde T=T/\ell$, 
 and $\tilde a_{\rm max}$ and  $\tilde a_{\rm min}$ 
are the maximum and minimum radii of the oscillating
universe.
We shall evaluate the period near the boundaries of
the parameter range
of oscillating universes (the light-orange region
in Fig. \ref{fig:Lambdap}(a)).
We first show the potential $\mathpzc{U}(\tilde a)$ 
for three (near-) boundary values of 
$\tilde g_{\rm s}$ in Fig. \ref{fig:pot_Lpkp}

For the case with
 an unstable static universe (the dashed blue curve)
($\Gamma_{1,1(+)}$ with $\tilde g_{\rm s}<0$),
the larger double root of the equation of $\mathpzc{U}(\tilde a)=0$ 
is given by
\begin{eqnarray}
\tilde a_S&=&\tilde a_S^{[1,1](+)}
:=\sqrt{
{1\over 3}\left(1+\sqrt{1-\tilde g_{\rm r}}\right)
}
\,,
\end{eqnarray}
while the smaller root is
\begin{eqnarray}
\tilde a_T&=&\tilde a_T^{[1,1](+)}
:=\sqrt{
{1\over 3}\left(1-2\sqrt{1-\tilde g_{\rm r}}\right)
}
\,,
\end{eqnarray}
which corresponds to
a turning radius at a bounce.
The period $T$ diverges in the limit of a static universe,
because $\tilde a_{\rm max}=\tilde a_S$ is the double root.

While, near a stable static universe (the solid blue curve)
($\Gamma_{1,1(-)}$), the period is finite and is evaluated as
\begin{eqnarray}
\tilde T_S&=&
\left({3\lambda-1\over 2}\right)^{1/2}\times
\pi
\left[1-(1-\tilde g_{\rm r})^{1/2}\over
 3(1-\tilde g_{\rm r})^{1/2}\right]^{1/2}
\,,~~~
\\
&\approx&
\left({3\lambda-1\over 2}\right)^{1/2}\times
\left \{
\begin{array}{ll}
\displaystyle
{\pi\over \sqrt{6}}\, \tilde g_{\rm r}^{1/2}
&(\tilde g_{\rm r}\ll 1) \\[1em]
\displaystyle
{\pi\over \sqrt{3}}\, {1\over (1-\tilde g_{\rm r})^{1/4}}
&(\tilde g_{\rm r}\approx 1)
\,.
\end{array}
\right.
~~~~~~\nonumber
\end{eqnarray}
The period $\tilde T_S$ changes from 0 to $\infty$ along the static curve
$\Gamma_{1,1(-)}$.

The radius of this stable static universe is given by
$\tilde a_S=\tilde a_S^{[1,1](-)}$, which is the smaller root of 
the equation of $\mathpzc{U}(\tilde a)=0$. 
The larger root $\tilde a_T=\tilde a_S^{[1,1](+)}$ 
corresponds to a turning radius 
of  a bounce universe, which is shown by $\tilde a_{T}$ in Fig. 
\ref{fig:pot_Lpkp}.

There is another boundary limit, i.e., $\tilde g_{\rm s}
\rightarrow 0^-$.
In this limit, we find the roots of $\mathpzc{U}(\tilde a)=0$ as
\begin{eqnarray}
\tilde a_{\rm 1}^2&\approx &0
\\
\tilde a_{\rm 2}^2&\approx 
&{1\over 2}\left(1-\sqrt{1-{4\over 3}\tilde g_{\rm r}}
\right)
\\
\tilde a_{\rm 3}^2&\approx 
&{1\over 3}\left(1+\sqrt{1-{4\over 3}\tilde g_{\rm r}}
\right)
\,.
\end{eqnarray}
Since the largest root ($\tilde a_{\rm 3}$) 
corresponds to a turning radius $\tilde a_{T}$ of 
 a bounce universe, the oscillation range is
$[\tilde a_1,\tilde a_2]$, and then 
 the period is evaluated approximately  by
\begin{eqnarray}
\tilde T_0&=&2\int_{0}^{\tilde a_{\rm 2}}
{d\tilde a\over \sqrt{-2\mathpzc{U}(\tilde a)}}
\,.
\label{period_nonzero_L2}
\end{eqnarray}
The period is then given by
\begin{eqnarray}
\tilde T_0&=&
\left({3\lambda-1\over 2}\right)^{1\over 2}\times
2
\sinh^{-1}\left[
{1-\left(1-{4\over 3}\tilde g_{\rm r}\right)^{1\over 2}\over
 2\left(1-{4\over 3}\tilde g_{\rm r}\right)^{1\over 2}}\right]^{1\over 2}
\,,~~~
\nonumber \\
&\approx&
\left({3\lambda-1\over 2}\right)^{1\over 2}\times
\left \{
\begin{array}{ll}
\displaystyle
{2\over \sqrt{3}}\, \tilde g_{\rm r}^{1\over 2}
&(\tilde g_{\rm r}\ll 1) \\[1em]
\displaystyle
 \ln \left[{\sqrt{3}\over 
 \left({3\over 4}-\tilde g_{\rm r}\right)^{1\over 2}}\right]
&(\tilde g_{\rm r}\approx {3\over 4})
\,.
\end{array}
\right.
~~~~~~
\end{eqnarray}
The period $\tilde T_0$ also changes from 0 to $\infty$ along the
 $\tilde g_{\rm r}$-axis;
$\tilde g_{\rm s}=0$ ($0<\tilde g_{\rm r}<3/4$).

We summarize our result as
$\tilde T\sim g_{\rm r}^{1/2}$ when $\tilde g_{\rm r}\ll 1$,
but it diverges near $\Gamma_{1,1(+)}$,
on which we have the unstable static universe.

\subsection{$\Lambda<0 ~(\epsilon=-1)$}
\label{Lambda-}

\begin{figure}[h]
\begin{center}
\includegraphics[scale=.4]{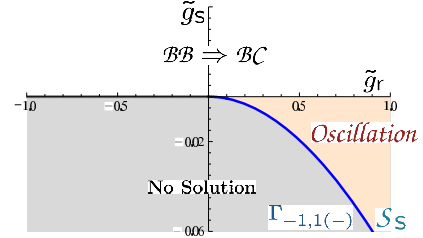}
\\
(a) $K=1$
\\[2em]
\includegraphics[scale=.4]{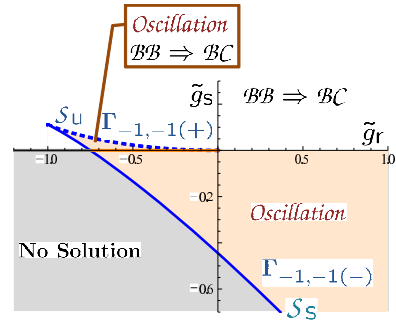}
\\
(b) $K=-1$
\caption{Phase diagram of spacetimes for $\Lambda<0$.
The oscillating universe is found for both  $K=\pm 1$.
The static universe exists on the boundary
$\Gamma_{-1,1(-)}$ ($K=1$) and
on  $\Gamma_{-1,-1(\pm)}$ ($K=-1$). 
In the branch of unstable static universe
on $\Gamma_{-1,-1(+)}$,
we also find dynamical
universes with an asymptotically static spacetime;
$\mathpzc{BB}$ $\Rightarrow$ $\mathpzc{S}_{\rm u}$,
$\mathpzc{S}_{\rm u}$ $\Rightarrow$ $\mathpzc{BC}$,
or 
$\mathpzc{S}_{\rm u}$ $\Rightarrow$ $\mathpzc{Bounce}$$\Rightarrow$
 $\mathpzc{S}_{\rm u}$.
}
\label{fig:Lambdam}
\end{center}
\end{figure}

The potential is 
given by 
\begin{eqnarray}
\mathpzc{U}(a)={1\over (3\lambda-1)\tilde a^4}\left[
K\tilde a^4+ \tilde a^6
-{\tilde g_{\rm r}\over 3}\tilde a^2-{\tilde g_{\rm s}
\over 3}\right]
\,,
\end{eqnarray}
We summarize our result in Fig. \ref{fig:Lambdam}.
\vskip -.2cm

In this case, if $\tilde g_{\rm s}>0$, we 
find a big bag  and a big crunch singularities
($\mathpzc{BB} \Rightarrow \mathpzc{BC}$) 
except for a small region in $K=-1$.
If $\tilde g_{\rm s}<0$, however,
we always find an oscillating universe
if the solution exists.

The conditions for an oscillating universe is shown by
the light-orange region in Fig. \ref{fig:Lambdam},
which is given by the following inequalities:\\
For $K=1$, 
\begin{eqnarray}
&&
\tilde g_{\rm r}>0
\nonumber \\[.5em]
&&
\tilde g_{\rm s}^{[-1,1](-)}(\tilde g_{\rm r})\leq
\tilde g_{\rm s}
<0
\,,
\end{eqnarray}
and for $K=-1$, 
\begin{eqnarray}
&
\tilde g_{\rm s}^{[-1,-1](-)}(\tilde g_{\rm r})\leq
\tilde g_{\rm s}
<0~~~~~~~~~~~~~~~~~
&~~{\rm with}~~\tilde g_{\rm r}\geq 0
\,,
\nonumber \\
&
\tilde g_{\rm s}^{[-1,-1](-)}(\tilde g_{\rm r})\leq
\tilde g_{\rm s}
\leq
\tilde g_{\rm s}^{[-1,-1](+)}(\tilde g_{\rm r})
&~~{\rm with}~~\tilde g_{\rm r}< 0
\,.
\nonumber 
\\
&&
~~
\end{eqnarray}
In the limit of $\tilde g_{\rm r}\ll 1$
(i.e. $\Lambda\rightarrow 0$) for $K=1$, we recover
the condition (\ref{cond_0}).

The boundary of the range of 
oscillating universe is given by 
the positive $\tilde g_{\rm r}$-axis, and
 $\Gamma_{-1, 1(-)}$ for $K=1$,
and $\Gamma_{-1, -1(\pm)}$
for $K=-1$.
On those boundaries $\Gamma_{-1, K(\pm)}$, 
which are defined by
$\tilde g_{\rm s}=\tilde g_{\rm s}^{[-1,1](-)}(\tilde g_{\rm r})$
($K=1$)
and $\tilde g_{\rm s}=\tilde g_{\rm s}^{[-1,-1](\pm)}
(\tilde g_{\rm r})$
($K=-1$),
we find a stable
and unstable static universes.

\begin{figure}[h]
\begin{center}
\includegraphics[scale=.4]{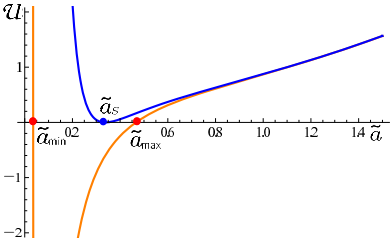}
\caption{
The potential 
$\mathpzc{U}(\tilde a)$  
for a stable static universe (blue)
 and an oscillating universe
 near $\tilde g_{\rm r}$-axis (orange)
in the case of $K=1$. We set
$\tilde g_{\rm r}= 0.8$ and $\tilde g_{\rm s}=-0.0477674$
 on $\Gamma_{-1,1}$ for a static
universe with the radius $\tilde a_S=0.337461$, and 
$\tilde g_{\rm r}= 0.8$ and $\tilde g_{\rm s}=-0.001$
for an oscillating universe, which maximum and minimum radii
are given by $\tilde a_{\rm max}=0.466615$ and 
$\tilde a_{\rm min}=0.035439$, respectively.
}
\label{fig:pot_Lmkp}
\end{center}
\end{figure}

\begin{figure}[h]
\begin{center}
\includegraphics[scale=.4]{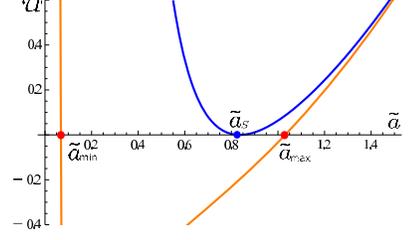}
\caption{
The potential 
$\mathpzc{U}(\tilde a)$ for a stable static universe (blue)
 and an oscillating universe
 near $\tilde g_{\rm r}$-axis (orange) for 
$K=-1$. We set 
$\tilde g_{\rm r}= 0.2$ and $\tilde g_{\rm s}=-0.581008$
 on $\Gamma_{-1,-1}$ for a static
universe, which radius is given by $\tilde a_S=0.835752$, and 
$\tilde g_{\rm r}= 0.2$ and $\tilde g_{\rm s}=-0.001$
for an oscillating universe, which maximum and minimum radii
are given by $\tilde a_{\rm max}=1.03075$ 
and $\tilde a_{\rm min}=0.0683656$, respectively.
}
\label{fig:pot_Lmkm1}
\end{center}
\end{figure}

\begin{figure}[h]
\begin{center}
\includegraphics[scale=.4]{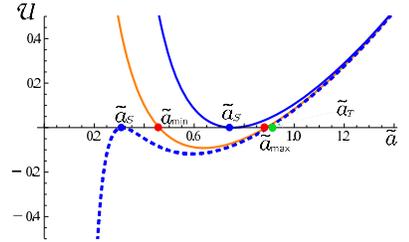}
\caption{
The potential 
$\mathpzc{U}(\tilde a)$ for a stable and unstable static universes
 (blue and red, respectively)
 and an oscillating universe
on $\tilde g_{\rm r}$-axis (dashed orange) for 
$K=-1$. We set
$\tilde g_{\rm r}= -0.5$ and $\tilde g_{\rm s}=-0.134123$
 on $\Gamma_{-1,-1(-)}$, and $\tilde g_{\rm s}=0.0230119$
 on $\Gamma_{-1,- 1(+)}$ for static
universes, which radius is given by $\tilde a_S=0.754344$, and 
$\tilde g_{\rm r}= -0.5$ and $\tilde g_{\rm s}=0$
for an oscillating universe, which maximum and minimum radii
are given by $\tilde a_{\rm max}=0.888074$
 and $\tilde a_{\rm min}=0.459701$, respectively.
We also find $\mathpzc{S}_{\rm u}$ $\Rightarrow$ $\mathpzc{Bounce}$
$\Rightarrow$ $\mathpzc{S}_{\rm u}$, which bounce radius is given by
$\tilde a_T=0.897072$. 
}
\label{fig:pot_Lmkm2}
\end{center}
\end{figure}

The period of an oscillating universe is given by 
Eq.(\ref{period_nonzero_L}).
We again evaluate its value near the boundary curves 
($\Gamma_{-1,K(-)}$)
and the positive $\tilde g_{\rm r}$-axis.
The potentials $\mathpzc{U}(\tilde a)$ for the (near-) boundary values of 
$\tilde g_{\rm s}$ are shown in Fig. \ref{fig:pot_Lmkp} ($K=1$), and
Figs.  \ref{fig:pot_Lmkm1} and \ref{fig:pot_Lmkm2} ($K=-1$). 


Note that the period diverges in the limit of an unstable
static universe (on $\Gamma_{-1,-1(+)}$),
where we find the radius of a static universe by
$\tilde a_S=\tilde a_{1}$
\begin{eqnarray} 
\tilde a_{1}^2:={1\over 3}\left(1-\sqrt{1+\tilde g_{\rm r}}\right)
\,.
\end{eqnarray}
The turning point is given by
$\tilde a_T=\tilde a_{2}$,
where
\begin{eqnarray} 
\tilde a_{2}^2:={1\over 3}\left(1+2\sqrt{1+\tilde g_{\rm r}}\right)
\,.
\end{eqnarray}

Near a stable static universe 
($\Gamma_{-1,1(-)}$ and $\Gamma_{-1,-1(-)}$), 
the period is evaluated as
\begin{eqnarray}
\tilde T_S&=&
\left({3\lambda-1\over 2}\right)^{1/2}\times
\pi
\left[(1+\tilde g_{\rm r})^{1/2}-K\over
 3(1+\tilde g_{\rm r})^{1/2}\right]^{1/2}
\,,~~~~~~
\end{eqnarray}
which approaches a constant
\begin{eqnarray}
\tilde T_S\approx 
{\pi\over \sqrt{3}}
\left({3\lambda-1\over 2}\right)^{1/2}
\label{period_asymptotic}
\end{eqnarray}
when $\tilde g_{\rm r}\gg 1$.
\begin{widetext}
Near the lower bound of $\tilde g_{\rm r}$, we find
\begin{eqnarray}
\tilde T_S&\approx&
{\pi\over \sqrt{3}}\left({3\lambda-1\over 2}\right)^{1/2}\times
\left \{
\begin{array}{lll}
\displaystyle
{\tilde g_{\rm r}^{1/2}\over\sqrt{2}} 
&\rightarrow 0
&({\rm as}~ \tilde g_{\rm r}\rightarrow 0~~{\rm for}~~K=1) \\[1em]
\displaystyle
(1+ \tilde g_{\rm r})^{-1/4}
&\rightarrow \infty
&({\rm as}~ \tilde g_{\rm r}\rightarrow -1~~{\rm for}~~K=-1) 
\,.
\end{array}
\right.
~~~~~~
\end{eqnarray}
\end{widetext}
Hence the period $T_S$ changes from 0 to a finite value 
(\ref{period_asymptotic}) 
along the curve
$\Gamma_{-1,1(-)}$ for $K=1$, while from
$\infty$ to at the same finite value along the curve
$\Gamma_{-1,-1(-)}$.

The radius of a static universe is given by
\begin{eqnarray}
\tilde a_S=\tilde a_S^{[-1,K](-)}=\sqrt{
{1\over 3}\left(\sqrt{1+\tilde g_{\rm r}}-K\right)
}
\,.
\end{eqnarray}
In the case of $\tilde g_{\rm r}<-3/4$ with $K=-1$,
there is another zero point of $\mathpzc{U}(\tilde a)$,
which gives a maximum turning point of 
 $\mathpzc{BB}$ $\Rightarrow$ $\mathpzc{BC}$, i.e.,
\begin{eqnarray}
\tilde a_{T}
=\tilde a_{T}^{[-1,-1](-)}
=\sqrt{
{1\over 3}\left(1-2\sqrt{1+\tilde g_{\rm r}}\right)
}
\,.
\end{eqnarray}

Near $\tilde g_{\rm r}$-axis,
 we find 
the solutions of the equation $\mathpzc{U}(\tilde a)=0$
as 
\begin{eqnarray}
\tilde a_\pm^2={1\over 2}\left(-K\pm\sqrt{1+{4\over 3}
\tilde g_{\rm r}}\right)
\,,
\end{eqnarray}
as well as $\tilde a_0\approx 0$.
We have a maximum radius $\tilde a_{\rm max}=\tilde a_+$,
and find that
the minimum radius $\tilde a_{\rm min}$ is 
almost zero for $\tilde g_{\rm r}>0$ because $\tilde a_-^2<0$, but 
in the case of $K=-1$, for  $-3/4<\tilde g_{\rm r}<0$,
we find a finite minimum radius 
$\tilde a_{\rm min}=\tilde a_-$.

Using those values, we evaluate the period as
\begin{eqnarray}
\tilde T_0&=&
\left({3\lambda-1\over 2}\right)^{1\over 2}\,
\sec^{-1}\sqrt{1+{4\over 3}\tilde g_{\rm r}}
\end{eqnarray}
for $K=1$, and
\begin{eqnarray}
\tilde T_0&=&
\left({3\lambda-1\over 2}\right)^{1\over 2}\times
\left \{
\begin{array}{l}
\displaystyle
\pi-\sec^{-1}\sqrt{1+{4\over 3}\tilde g_{\rm r}}
~~~
\left(\tilde g_{\rm r}\geq 0\right) 
\nonumber
\\[1em]
\displaystyle
~~~~~
\pi
~~~~~~~~~~~
\left(-{3/4}<\tilde g_{\rm r}<0\right)
\end{array}
\right.
~~~~~~
\end{eqnarray}
for $K=-1$. 
The period $\tilde T_0$ also changes from 0 to $\infty$ along the
 $\tilde g_{\rm r}$-axis;
$\tilde g_{\rm s}=0$ ($0<\tilde g_{\rm r}<3/4$).

In the case with  the detailed balance condition, 
since $\Lambda<0, \tilde g_{\rm r}=-9/4,  \tilde g_{\rm s}=0$,
we do not find any FLRW solution.
If we include matter fluid, the result will change.
For example, if we have ``radiation" fluid, which 
energy density is proportional to $a^{-4}$,
we should shift the value of  $\tilde g_{\rm r}$.
Then if $-3/4\leq \tilde g_{\rm r}<0$,
we 
find an oscillating universe 
 for $K=-1$,
which period is $\pi[(3\lambda-1)/2]^{1/2}$.
The equality ($\tilde g_{\rm r}=-3/4$)
 gives a static universe.

\section{Toward More Realistic Cosmological Model}
\label{Realistic Cosmology}
In the Ho\v{r}ava-Lifshitz gravity without
the detailed balance condition, we find a variety of phase structures
 of vacuum spacetimes depending on the coupling constants $g_{\rm r}$ and 
$g_{\rm s}$
as well as the spatial curvature $K$ and a cosmological constant
 $\Lambda$.
Note that there is no vacuum FLRW solution in the case with 
the detailed balance condition.
We summarize our result in Table \ref{table1}.
We have obtained an oscillating spacetime as well 
as a bounce universe for a wide range of coupling constants.
We have also evaluated the period of the oscillating universe.

\begin{center}
\begin{table}[h]
\footnotesize{
\begin{tabular}{|c||c|l|c|l|}
\hline
&\multicolumn{2}{c|}{$K=1$}
&\multicolumn{2}{c|}{$K=-1$}
\\
\hline
\hline
&\multicolumn{2}{l|}{$\ast$\,
\red{$\mathpzc{Oscillation}$}}
&\multicolumn{2}{l|}{~}
\\ 
&\multicolumn{2}{l|}{
$\ast$\,
\green{$\mathpzc{dS}$$ \Longleftrightarrow $$\mathpzc{Bounce}$}}
&\multicolumn{2}{l|}{
$\ast$\,
\green{$\mathpzc{dS}$$ \Longleftrightarrow $$\mathpzc{Bounce}$}}
\\
&\multicolumn{2}{l|}{$\ast$\,
$\mathpzc{BB}$
$\Rightarrow$$\mathpzc{BC}$}
&\multicolumn{2}{l|}{$\ast$\,
$\mathpzc{BB}$
$\Rightarrow$$\mathpzc{BC}$}
\\
&\multicolumn{2}{l|}{$\ast$\,
$\mathpzc{BB}\Rightarrow \mathpzc{dS}$ 
($\mathpzc{dS}\Rightarrow \mathpzc{BC}$)}
&\multicolumn{2}{l|}{$\ast$\,
$\mathpzc{BB}\Rightarrow \mathpzc{dS}$
($\mathpzc{dS}\Rightarrow \mathpzc{BC}$)}
\\[.2em]
\cline{2-5}
$\Lambda>0$
&{\tiny\blue{$\Gamma_{1,1(\pm)}$}}&~$\ast$\,
\blue{$\mathpzc{S}_{\rm u}$, 
$\mathpzc{S}_{\rm s}$}
&{\tiny\blue{$\Gamma_{1,-1(+)}$}}& ~$\ast$\,
\blue{$\mathpzc{S}_{\rm u}$}
\\[.2em]
\cline{2-2}\cline{4-4}
&\multicolumn{2}{l|}{$\ast$\,
$\mathpzc{BB}$$\Rightarrow$$\mathpzc{S}_{\rm u}$
($\mathpzc{S}_{\rm u}$ $\Rightarrow$ $\mathpzc{BC}$) }
&\multicolumn{2}{l|}{$\ast$\,
$\mathpzc{BB}$$\Rightarrow$$\mathpzc{S}_{\rm u}$
($\mathpzc{S}_{\rm u}$ $\Rightarrow$ $\mathpzc{BC}$) }
\\
&\multicolumn{2}{l|}{
$\ast$\,
\green{$\mathpzc{S}_{\rm u}$ $\Rightarrow$ $\mathpzc{dS}$
($\mathpzc{dS}$$\Rightarrow$ $\mathpzc{S}_{\rm u}$)}}
&\multicolumn{2}{l|}{$\ast$\,
\green{$\mathpzc{S}_{\rm u}$ $\Rightarrow$ $\mathpzc{dS}$
($\mathpzc{dS}$$\Rightarrow$ $\mathpzc{S}_{\rm u}$)}}
\\
&\multicolumn{2}{l|}{$\ast$\,
\red{$\mathpzc{S}_{\rm u}$ $ \Longleftrightarrow $ $\mathpzc{Bounce}$
} }
&\multicolumn{2}{l|}{~}
\\[.2em]
\hline
&\multicolumn{2}{l|}{$\ast$\,
\red{$\mathpzc{Oscillation}$}}
&\multicolumn{2}{l|}{$\ast$\,
\green{$\mathpzc{M}$$ \Longleftrightarrow $$\mathpzc{Bounce}$}}
\\
&\multicolumn{2}{l|}{$\ast$\,
$\mathpzc{BB}$
$\Rightarrow$$\mathpzc{BC}$}
&\multicolumn{2}{l|}{$\ast$\,
$\mathpzc{BB}$
$\Rightarrow$$\mathpzc{BC}$}
\\
$\Lambda=0$
&\multicolumn{2}{l|}{~}
&\multicolumn{2}{l|}{$\ast$\,
$\mathpzc{BB}\Rightarrow \mathpzc{M}$
($\mathpzc{M}\Rightarrow \mathpzc{BC}$)}
\\[.2em]
\cline{2-5}
&{\tiny\blue{$\Gamma_{0,1}$}}&
&{\tiny\blue{$\Gamma_{0,-1}$}}&
~$\ast$\,
\blue{$\mathpzc{S}_u$}
\\[.2em]
\cline{2-2}\cline{4-4}
&
\multicolumn{2}{l|}{$\ast$\,
\blue{$\mathpzc{S}_s$}
}
& 
\multicolumn{2}{l|}{$\ast$\,
$\mathpzc{BB}\Rightarrow \mathpzc{S}_u$
($\mathpzc{S}_u \Rightarrow \mathpzc{BC}$)}
\\
&\multicolumn{2}{l|}{~}
&\multicolumn{2}{l|}{$\ast$\,
\green{$\mathpzc{S}_u \Rightarrow \mathpzc{M}$
($\mathpzc{M}\Rightarrow \mathpzc{S}_u$)}}
\\[.2em]
 \hline
&\multicolumn{2}{l|}{$\ast$\,
\red{$\mathpzc{Oscillation}$}}
&\multicolumn{2}{l|}{$\ast$\,\red{$\mathpzc{Oscillation}$}}
\\
&\multicolumn{2}{l|}{$\ast$\,
$\mathpzc{BB}$
$\Rightarrow$$\mathpzc{BC}$}
&\multicolumn{2}{l|}{$\ast$\,
$\mathpzc{BB}$
$\Rightarrow$$\mathpzc{BC}$}
\\[.2em]
\cline{2-5}
$\Lambda<0$
&{\tiny\blue{$\Gamma_{-1,1(-)}$}}&
&{\tiny\blue{$\Gamma_{-1,-1(\pm)}$}}&
~$\ast$\,
\blue{$\mathpzc{S}_u$, $\mathpzc{S}_s$}
\\[.2em]
\cline{2-2}\cline{4-4}
&\multicolumn{2}{l|}{$\ast$\,
\blue{$\mathpzc{S}_s$}}
&
\multicolumn{2}{l|}{$\ast$\,
$\mathpzc{BB}\Rightarrow \mathpzc{S}_u$
($\mathpzc{S}_u \Rightarrow \mathpzc{BC}$)}
\\
&\multicolumn{2}{l|}{~}
&\multicolumn{2}{l|}{$\ast$\,
\red{$\mathpzc{S}_{\rm u}$ $ \Longleftrightarrow $ $\mathpzc{Bounce}$
} }
\\
 \hline
\end{tabular} }
\caption{
Summary: What type of
spacetime is possible for each $\Lambda$ and each $K$.
Non-singular universes are shown by the colored letters
(an oscillating universe and dynamical spacetimes 
evolving in a finite scale range 
by red, static universes by blue, 
dynamical spacetimes  evolving from or to 
an  asymptotically infinite scale 
by green).
$\mathpzc{dS}$, $\mathpzc{BB}$, $\mathpzc{BC}$, 
$\mathpzc{S}_{\rm u}$, $\mathpzc{S}_{\rm s}$
and $\mathpzc{M}$ denote de Sitter space, a big bang,
a big crunch, an unstable static universe,
a stable static universe, and Milne universe, respectively.
}
\label{table1}
\end{table}
\end{center}
In our analysis, we assume that 
the integration constant $C$ from the projectability condition
vanishes.
If $C\neq 0$, one may find a different story.
In fact, if $g_{\rm s}=0$ and $g_{\rm r}<0$
just as the case with the detailed balance condition,
we will find the similar vacuum solutions to 
 the present ones, because
$C$ and $g_{\rm r}$ without $g_{\rm s}$-term
 play the similar roles to
those of $g_{\rm r}$ and $g_{\rm s}$ in the present model.
For example, we obtain 
an oscillating universe for large $C(>0)$
with $g_{\rm s}=0$, $g_{\rm r}<0$, $\Lambda=0$ and $K=1$.
This avoidance of a singularity is, however,  caused by
the negative ``radiation" density  from the higher curvature terms.
Hence if one includes the conventional radiation,
then the effective $g_{\rm r}$ becomes positive
as we will show below, and as a result
the universe will inevitably collapse to a big-crunch singularity.
Furthermore, if radiation field 
evolves as $a^{-6} $ in the UV limit\cite{cosmology3},
the inclusion of such radiation  will 
kill the possibility of singularity avoidance
by ``dark" radiation.

As we have evaluated,
 the oscillation period and amplitude are expected to be 
the Planck scale or the scale $\ell$ defined by a cosmological constant
$\Lambda$, unless the coupling constants are unnaturally large.
Hence it cannot be a cyclic universe, which period is 
macroscopic such as  the age of the universe.

In order to find more realistic universe,
we have to include some other components, 
which we shall discuss here.
First of all, one may claim inclusion of matter fluid.
When we include a dust fluid ($P=0$),
the conventional radiation  ($P=\rho/3$), and stiff matter
 ($P=\rho$),
we can treat such a case just by replacing the constant $g_{\rm d}$,
$g_{\rm r}$ and $g_{\rm s}$ with
\begin{eqnarray}
g_{\rm d}&=&8C+g_{\rm dust}
\nonumber \\
g_{\rm r}&=&6(g_3+3g_2)+g_{\rm rad}
\nonumber \\
g_{\rm s}&=&12(9g_5+3g_6+g_7)K+g_{\rm stiff}
\,,
\end{eqnarray}
 where $g_{\rm dust}$, $g_{\rm rad}$ and 
$g_{\rm stiff}$, which come from real dust fluid,
radiation and stiff matter, are positive constants.
In this case, the present analysis is still valid.
If $g_{\rm rad}$ is large enough just as our universe,
a maximum scalar factor $a_{\rm max}$ 
of the the oscillating universe will become large
(see, for example,  Eq. (\ref{aminmax})), and then
it can be  a cyclic universe.

If the equation of state is still given by
$P=w\rho$ ($w$=constant), the analysis is straightforward.
When we have other types of matter fields,
e.g. a scalar field with a potential,
the analysis will be more complicated.
The phase space analysis may be appropriate
 for the case with a scalar field
\cite{Halliwell}.

\begin{figure}[h]
\begin{center}
\includegraphics[scale=.4]{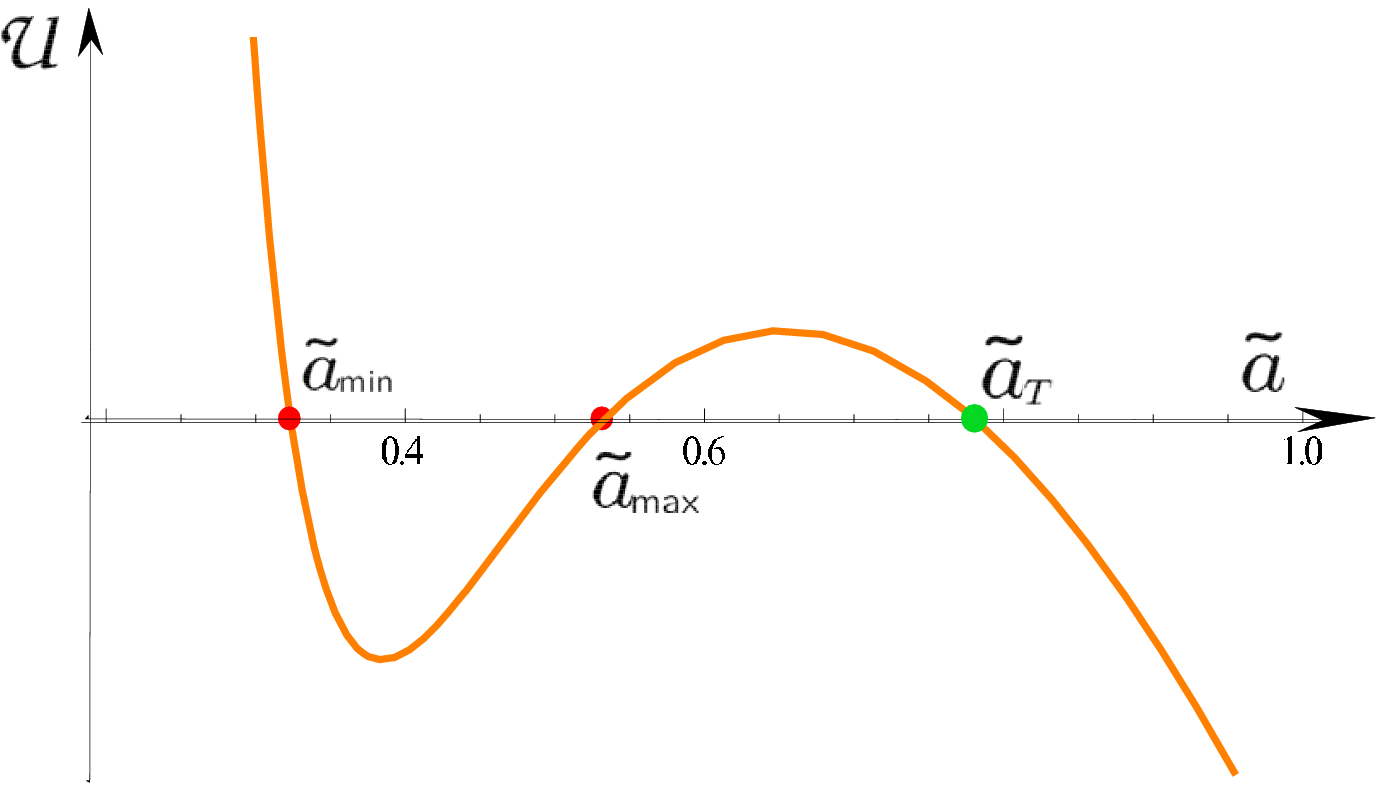}
\caption{
}
\label{fig:pot_LpKp_tunnel}
\end{center}
\end{figure}

From our present analysis, one may speculate the following 
``realistic" scenario for the early stage of the universe.
Suppose  a closed universe is created from ``nothing"
initially  in an oscillating phase
(see Fig. \ref{fig:pot_LpKp_tunnel})
\cite{Vilenkin-Hartle-Hawking,cosmology48}.
Such a universe may be very small and oscillating between 
two radii ($a_{\rm min}$ and $a_{\rm max}$) 
with a time scale $\ell$.
If  we have a positive cosmological constant ($\Lambda>0$),
there exists a potential barrier as shown in Fig.
 \ref{fig:pot_LpKp_tunnel}.

After numbers of oscillations, the universe may 
quantum mechanically tunnel to a bounce point $a_{T}$.
Then the universe will expand to de Sitter phase
because a positive cosmological constant, finding
the  universe in a macroscopic scale\footnote{
After we have written up 
this paper, we have found \cite{emergent_universe}, in which 
a cosmological transition scenario from 
a static (or an oscillating) universe to an inflationary stage
was discussed. They assume that
the equation of state changes in time, which mechanism is 
not specified.}.
Furthermore, one can refine this scenario, if 
 there exists a scalar field, which is responsible for inflation,
instead of a cosmological constant.
Before tunneling, we may find the similar scenario to the above one.
After tunneling, 
the potential of the scalar field will behaves as
a cosmological constant in  a slow-rolling period.
We will find an exponential expansion of the universe
after tunneling.
However, inflation will eventually end and 
the energy of the scalar field is converted to that of
conventional matter fluid via a reheating of the universe.
We find a big bang universe. Since the universe is closed,
but the scale factor has lower bound because of negative 
``stiff matter", we will find 
a macroscopically large cyclic universe after all.
To confirm such a scenario, we should analyze the dynamics
of the universe with an inflaton field in detail.
The work is in progress.

We also have another extension of the present
FLRW spacetime to anisotropic one.
It may be interesting and important 
not only to study the dynamics of Bianchi spacetime
\cite{cosmology20,cosmology21}
but also to analyze the stability of the FLRW universe
against anisotropic perturbations\cite{future}.
\acknowledgments

We would like to thank Yuko Urakawa
for valuable  comments and discussions.
This work was partially supported 
by the Grant-in-Aid for Scientific Research
Fund of the JSPS (No.22540291) and for the
Japan-U.K. Research Cooperative Program,
and by the Waseda University Grants for Special Research Projects. 

\appendix
\section{stability of a flat background and the coupling constants}

In this Appendix, we discuss the conditions on
the coupling constants by which gravitons are perturbatively stable.
From the perturbation analysis around a flat background,
we obtain the dispersion relation for
the usual helicity-2 polarizations of the graviton~\cite{WangM},
\begin{eqnarray}
\omega_{\rm TT(\pm)}^2=-g_1k^2+g_3{k^4\over M_{\rm PL}^2}
\pm g_4{k^5\over M_{\rm PL}^3}+g_9{k^6\over M_{\rm PL}^4}.
\end{eqnarray}
The stability both in the IR and UV regimes requires
\begin{eqnarray}
g_1<0,
\quad
g_9>0.
\end{eqnarray}
By a suitable rescaling of time, we then set $g_1=-1$.

As a result of the reduced symmetry~(\ref{f-p})
the longitudinal degree of freedom of the graviton appears,
and its stability is more subtle.
First of all the longitudinal graviton is plagued with ghost instabilities
for $1/3<\lambda<1$~\cite{Horava}.
The dispersion relation for the longitudinal mode turns out to be~\cite{WangM}
\begin{eqnarray}
\left({3\lambda-1\over \lambda-1}\right)\omega_{\rm L}^2
&=&g_1k^2+(8g_2+3g_3){k^4\over M_{\rm PL}^2}
\nonumber \\&&
+(-8g_8+3g_9){k^6\over M_{\rm PL}^4}
\,.
\end{eqnarray}
We see that the sound speed squared is negative in the IR if $g_1<0$ and
 $\lambda>1$,
which implies that the longitudinal graviton is unstable in the IR~\cite{K-A}.
However, this fact itself does not necessarily mean that the theory 
suffers from pathologies,
because whether or not an instability really causes a trouble depends upon 
its time scale~\cite{Izumi}.
Moreover, there is an attempt to improve the behavior of the longitudinal 
graviton
by promoting $N$ to an $\Vec{x}$-dependent function and
adding terms constructed from the 3-vector $\partial_i N/N$ in the 
Lagrangian~\cite{BPS_2}.\footnote{Obviously,
in this case the Hamiltonian constraint is imposed locally and
the additional dust-like component does not appear in the Friedmann equation.}
It can be shown that
the non-projectable Ho\v{r}ava gravity thus extended appropriately does not 
plagued with
instabilities of the longitudinal gravitons~\cite{BPS_2}.
In light of these subtleties, we do not consider the stability of the 
longitudinal sector furthermore,
while we do require the stability for the usual helicity-2 polarizations of the
graviton.

Note that the detailed balance condition satisfies 
$g_1<0$ and $g_9>0$.

\section{quantum tunneling from an oscillating universe}
In the case of $K=1$ and $\Lambda>0$, we have a bouncing universe as
well as an oscillating universe. These two solutions 
are separated by a finite potential wall as we see 
in Fig \ref{fig:pot_LpKp_tunnel}. 
Hence we expect quantum tunneling from an oscillating universe
to an exponentially expanding universe.
In this Appendix, we shall evaluate the tunneling probability.

First we consider the normalized Euclidean metric 
\begin{eqnarray}
d\tilde s^2=d\tilde \tau^2+\tilde b^2(\tilde \tau)d\Sigma_{K=1}^2
\,,
\end{eqnarray}
which satisfies the following equation
\begin{eqnarray}
\tilde b'^2-2\mathpzc{U}(\tilde b)=0
\,,
\label{F_eq_Euclid}
\end{eqnarray}
where the prime denotes the derivative with respect to the Euclidean time 
$\tilde \tau$,
and the potential $\mathpzc{U}$ is written as
\begin{eqnarray}
2\mathpzc{U}(\tilde b)={2\over 3\lambda-2}{1\over \tilde b^4}\left[
-(\tilde b^2-\tilde b_{\rm max}^2)(\tilde b^2-\tilde b_{\rm min}^2)
(\tilde b^2-\tilde b_T^2)
\right]
\,.
\end{eqnarray}
The variables with a tilde are normalized ones by use of
the scale length $\ell=\sqrt{3/\Lambda}$ just as in the text.
The bounce solution $\tilde b(\tilde \tau)$ is obtained by integraton of Eq.
(\ref{F_eq_Euclid}).
The Euclidean action is given by
\begin{eqnarray}
S_E=3(3\lambda-1)\ell \int d\tilde \tau d^3x \tilde b\left[{1\over 2}
\tilde b'^2+\mathpzc{U}(\tilde b)\right]
\,.
\end{eqnarray}
Using Eq. (\ref{F_eq_Euclid}), we find the action $S_E$ as
\begin{eqnarray}
S_E=3(3\lambda-1)\ell^2 V_3 \int d\tilde b \tilde b  
\sqrt{2\mathpzc{U}(\tilde b)}
\,,
\end{eqnarray}
where $V_3=2\pi^2$ is the volume of a unit three sphere.
Introducing $u$ by
\begin{eqnarray}
\tilde b^2=\tilde b_T^2(1-k^2 u^2)\,,
\end{eqnarray}
where
$k^2=(\tilde b_T^2-\tilde b_{\rm max}^2)/\tilde b_T^2
(<1)$.
We then find
\begin{eqnarray}
S_E&=&{12\pi^2 \ell^2 \over \kappa^2}
{(\tilde b_T^2-\tilde b_{\rm max}^2)^2
(\tilde b_T^2-\tilde b_{\rm min}^2)^{1/2}}
\nonumber \\
 &\times&
\int_0^1 {u^2du\over 1-k^2u^2}
\sqrt{(1-u^2)(1-m^2u^2)}
\,,
\end{eqnarray}
where $m^2=(\tilde b_T^2-\tilde b_{\rm max}^2)/
(\tilde b_T^2-\tilde b_{\rm min}^2) (<1)$.

It can be easily evaluated in the limit of a static universe, i.e.,
$\tilde g_{\rm s}=\tilde g_{\rm s}^{[1,1](-)}(\tilde g_{\rm r})$.
Using 
 $\tilde b_{\rm max}\approx \tilde b_{\rm min}\approx \tilde b_S$,
we find
\begin{eqnarray}
S_E&=&{12\pi^2 \ell^2  \over \kappa^2}
(\tilde b_T^2-\tilde b_S^2)^{5/2}
\nonumber \\
&\times&
\left({3-2k^2\over 3k^4}-{1-k^2\over k^5}\tanh^{-1}k
\right)\,,
\end{eqnarray}
where 
$k=\sqrt{\tilde b_T^2-\tilde b_S^2}/\tilde b_T$.
Since $\tilde b_T^2=(1+2\sqrt{1-\tilde g_{\rm r}})/3$
and $\tilde b_T^2-\tilde b_S^2=\sqrt{1-\tilde g_{\rm r}}$,
we find 
\begin{eqnarray}
S_E&=&{4\pi^2 \ell^2 \over \kappa^2} (1-\tilde g_{\rm r})^{1/4}
\nonumber \\
&\times&
\left[1-{(1+2\sqrt{1-\tilde g_{\rm r}})^{1/2}
(1-\sqrt{1-\tilde g_{\rm r}})\over \sqrt{3}(1-\tilde g_{\rm r})^{1/4}}
\tanh^{-1}k
\right]\,,
\nonumber \\
~~
\end{eqnarray}
with
\begin{eqnarray}
k^2={3\sqrt{1-\tilde g_{\rm r}}\over 1+2\sqrt{1-\tilde g_{\rm r}}}
\,.
\end{eqnarray}
The tunneling probability is given by
$P\sim e^{-{S_E}}$.

\begin{figure}[h]
\begin{center}
\includegraphics[scale=.4]{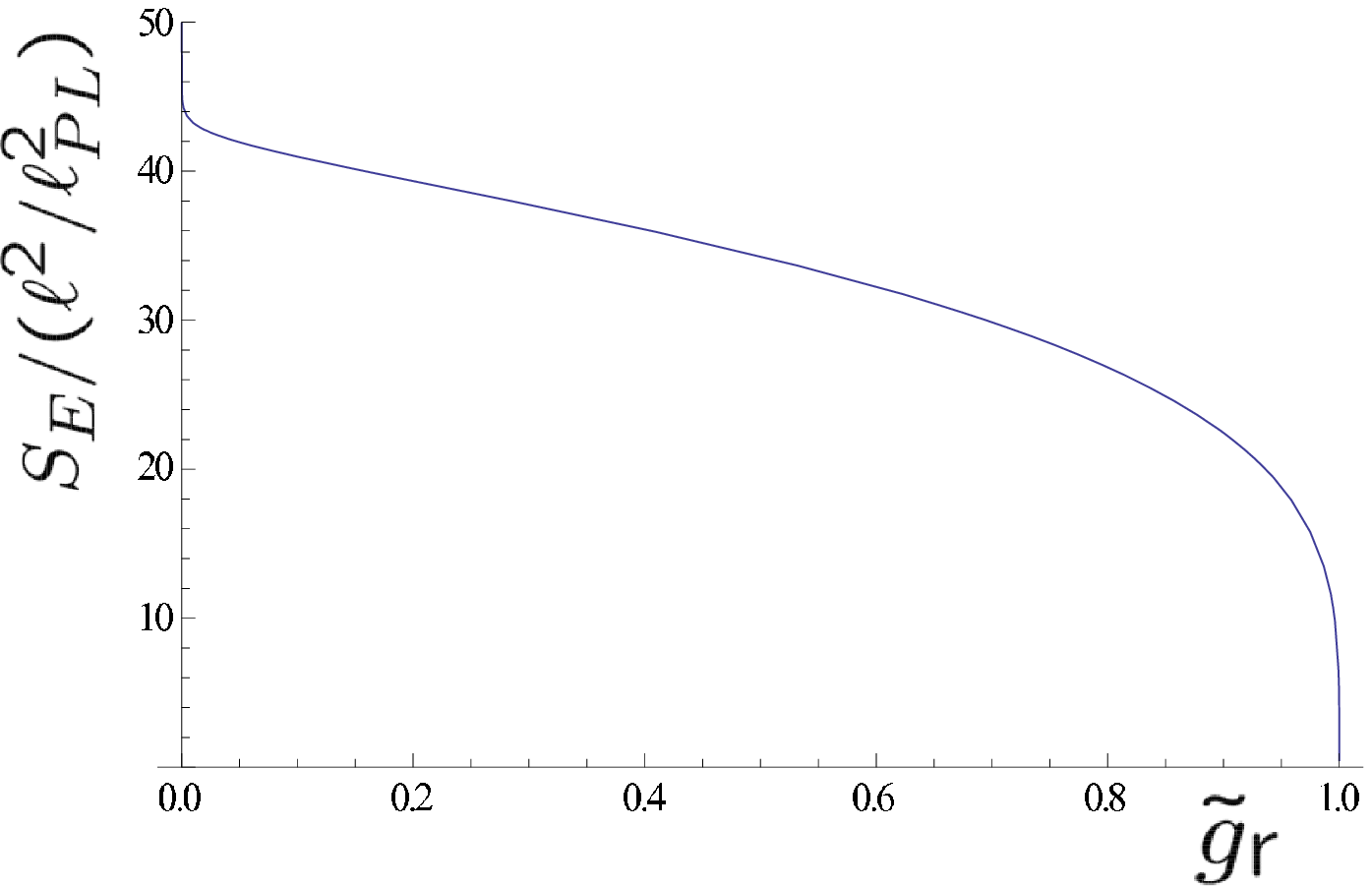}
\caption{
}
\label{S_E}
\end{center}
\end{figure}

We show the behavior of $S_E$ in Fig. \ref{S_E}.
We find 
\begin{eqnarray}
P&\sim& \exp\left[{-(20-40)\times \left({\ell\over \ell_{\rm PL}}\right)^2}
\right]
\nonumber \\
&\sim &
\exp\left[-(60-120)\times \left({m_{\rm PL}^4\over \rho_{\rm vac}}\right)
\right]
\end{eqnarray}
except for two limiting cases: $\tilde g_{\rm r}\sim 1$,
in which $S_E$ vanishes, and $\tilde g_{\rm r}\sim 0$, in which
$S_E$ diverges. In the former case, the potential barrier vanishes
giving a high tunneling probability, while in the latter case,
the potential barrier diverges giving zero tunneling probability.
If the vacuum energy (or potential) just after tunneling is the Planck scale,
the probability is evaluated as $P\sim e^{-(60-120)}$, which is very small
but finite.

\end{document}